# PP-LEM: Efficient and Privacy-Preserving Clearance Mechanism for Local Energy Markets ☆

Kamil Erdayandi [a],[*], Mustafa A. Mustafa [a],[b]

[a] *Department of Computer Science, The University of Manchester, Oxford Road, Manchester, M13 9PL, United Kingdom*
[b] *COSIC, KU Leuven, Leuven, 3001, Belgium*

ARTICLE INFO

ABSTRACT

In this paper, we propose a novel Privacy-Preserving clearance mechanism for Local Energy Markets (PP-LEM), designed for computational efficiency and social welfare. PP-LEM incorporates a novel *competitive* game-theoretical clearance mechanism, modelled as a Stackelberg Game. Based on this mechanism, a privacy-preserving market model is developed using a partially homomorphic cryptosystem, allowing buyers' reaction function calculations to be executed over encrypted data without exposing sensitive information of both buyers and sellers. The comprehensive performance evaluation demonstrates that PP-LEM is highly effective in delivering an incentive clearance mechanism with computational efficiency, enabling it to clear the market for 200 users within the order of seconds while concurrently protecting user privacy. Compared to the state of the art, PP-LEM achieves improved computational efficiency without compromising social welfare while still providing user privacy protection.

## 1. Introduction

The usage of traditional fossil fuel-based energy sources for generating power, such as coal and fuel oil, is gradually being phased out owing to their negative environmental effect [1–3]. In accordance, there is a trend towards using environmentally-friendly alternatives — Renewable Energy Sources (RES) such as wind and solar [4]. Yet, while the use of RES provides a chance to reduce environmental damage, it also introduces new challenges to our grid that must be handled. One of the major challenges with RES is their inherent instability as energy sources. In contrast to traditional sources, the output of RES fluctuates with weather conditions, creating unpredictability in power supply [5]. As Liang et al. [6] point out that this unpredictability makes grid balancing more complex and less efficient. This difficulty is worsened by the already volatile demand side of the power market.

In addition, traditional electricity markets are not designed to deal with high penetration levels of RES. These markets rely on two-tiered retail pricing for buyers and low or zero Feed-in-Tariffs (FiTs) for sellers [7]. These markets do not account for the fluctuation of RES and are insufficient for encouraging the use of RES. As a result, despite the trend towards renewable energy, non-renewable fuels continue to be major energy sources for power generation in most places worldwide. Thus, It is critical to establish a market that allows local electricity producers to sell their excess energy at a higher price than FiTs and consumers to buy their required electricity from local RES owners at a lower price than retail prices to promote the use of RES [8,9].

In response to these challenges, Local Energy Markets (LEMs) have gained traction as a strategic solution to enhance the integration of RES into the energy grid [10,11]. LEMs aim to motivate individuals and small-scale entities acting as consumers, producers, or both (prosumers) to partake in energy exchanges within a competitive marketplace. This initiative seeks to locally align energy supply with demand [12]. With LEMs, it is now possible to establish a trading mechanism that allows RES owners to maximise their profits and reduce their bills by trading electricity directly with other users for a price higher than FiT or buying electricity for a price lower than retail prices [13,14].

However, LEM's widespread implementation is impeded by a variety of challenges. One of the difficulties is the possible privacy risks involved with the operation of LEMs [15], as LEMs require collecting, processing, and exchanging user data [16]. Access to this data creates significant risks to users' privacy [17,18]. Utilising user data in LEMs can inadvertently reveal personal energy habits and lifestyles. Analysing offers, bids, and consumption patterns can expose private details, such as a user's presence at home and their appliance usage. This connection between energy transactions and personal data presents a

☆ This work was supported in part by the EPSRC through the project EnnCore EP/T026995/1 and by the Flemish Government through the FWO SBO project SNIPPET S007619. K.E is funded by The Ministry of National Education, Republic of Türkiye.
* Corresponding author.
*E-mail addresses:* kamil.erdayandi@manchester.ac.uk, kamil.erdayandi@meb.k12.tr (K. Erdayandi), mustafa.mustafa@manchester.ac.uk (M.A. Mustafa).

https://doi.org/10.1016/j.segan.2024.101477
Received 7 March 2024; Received in revised form 3 July 2024; Accepted 16 July 2024
Available online 22 July 2024

**Nomenclature**

**Abbreviations**

| | |
|---|---|
| DP | Differential Privacy |
| EV | Electrical Vehicle |
| FHE | Fully Homomorphic Encryption |
| FiTs | Feed-in-Tariffs |
| GDPR | General Data Protection Regulation |
| HE | Homomorphic Encryption |
| MGCC | MicroGrid Centre Controller |
| MPC | Multi-Party Computation |
| NE | Nash Equilibrium |
| P2P | Peer-to-Peer |
| PFET | Privacy-Friendly Energy Trading |
| PHE | Partially Homomorphic Encryption |
| PP-LEM | Privacy-Preserving Clearance Mechanism for Local Energy Markets |
| PV | Photovoltaic |
| PVGIS | Photovoltaic Geographical Information System |
| RES | Renewable Energy Sources |
| SC | Settlement Cycle |
| SM | Smart Meter |
| TTP | Trusted Third Party |
| VCG | Vickrey–Clarke–Groves |

**Notations/Symbols**

| | |
|---|---|
| $\epsilon$ | Criteria for stopping the iterations based on maximum demand–supply discrepancy |
| $\eta_1$ | Price adjustment fraction value for the next iteration |
| $\{.\}_{\mathcal{E}}$ | Homomorphically encrypted data |
| $\lambda_i$ | Profile variable of $b_i$, linked to prosumers' behaviours |
| $\overline{W}$ | Average welfare |
| $\pi_j$ | Price of electricity proposed by $s_j$ |
| $\rho_{FiT}$ | Feed-in-Tariffs (FiTs) price |
| $\rho_{Sup}$ | Supplier price |
| $\hat{D}_{Tot}$ | Final total demand |
| $\mathbb{Z}$ | Set of integers |
| $\mathbb{Z}_n^*$ | Set of integers relatively prime to n |
| $\mathbb{Z}_{n^2}^*$ | Set of integers relatively prime to n² |
| $b_i$ | $i$th buyer |
| $D_j$ | Total demand on $s_j$ |
| $D_{Tot}$ | Total demand from buyers |
| $lcm(.)$ | Least common multiple function |
| $N_B$ | Number of buyers |
| $N_S$ | Number of sellers |
| $N_{Iter}$ | Number of iterations |
| $PK_S$ | Public key of sellers |
| $S_j^{P2P}$ | Amount of electricity that $s_j$ trade in P2P market |
| $S_j$ | Amount of electricity that $s_j$ can supply |
| $s_j$ | $j$th seller |
| $S_{Tot}$ | Total supply from sellers |
| $SK_S$ | Private key of sellers |
| $t_k$ | $k$th trading period |
| $W_{B_j}$ | Welfare, payoffs of all buyers obtained from $s_j$ |
| $X_{ji}$ | Amount of electricity that $b_i$ wishes to buy from $s_j$ |

**Units**

| | |
|---|---|
| kWh | Kilo-watt hours |
| s | Seconds |

crucial to protect against unauthorised access [24]. Implementing a privacy-preserving system becomes essential to encourage the adoption of trading platforms in smart grids. Such a system can serve as a key incentive for utilising trading platforms, thereby encouraging RES adoption [8,9].

Computational efficiency is another critical aspect that needs to be considered in LEM models [25]. This can be achieved by implementing efficient algorithms that can handle large volumes of data generated by intelligent devices [26]. In addition, the system should be designed to minimise the overhead associated with data processing. Designing a computationally efficient system can significantly reduce latency in data processing activities [27]. Specifically, the energy trading in LEMs occurs in trading periods [28]; if computational tasks can be executed more swiftly with reduced latency, a larger number of users could engage in trading within a single trading slot, thereby maximising the market's capacity and efficiency.

Moreover, offering economic incentives is essential to motivate user participation in LEMs. A practical approach to achieve this is through the adoption of a pricing mechanism that accurately reflects the supply and demand dynamics within LEM. Such a strategy ensures that prices are equitable and truly representative of the energy's value, thereby promoting social welfare among users [29]. Thus, the pursuit of social welfare, which serves as a key incentive, should not be compromised by the concurrent objectives of ensuring privacy and computational efficiency.

In summary, to ensure the success of energy trading models in LEMs, it is crucial to have a secure, privacy-preserving, and computationally efficient system that provides economic incentives for users.

Within LEM models, critical activities include the generation of bids, market clearance, and the implementation of billing and settlement mechanisms. Billing and settlement mechanisms involve payment and transaction monitoring [28,30,31]. In contrast, the market clearance mechanism with initial bids is concerned with determining the price and quantity of energy to be exchanged between users, ensuring a match between supply and demand [15,32]. In this study, our emphasis lies in introducing an efficient mechanism for preserving privacy in market clearance. Therefore, we further explore the existing works on privacy-preserving market clearance mechanisms. These techniques [8, 25,33–49] aim to enhance market clearance in trading activities by focusing on security, privacy, and efficiency. However, there is no one-size-fits-all approach for addressing all aspects of efficiency, security, and privacy in clearance mechanisms. The effectiveness of these measures may vary based on the unique risks and characteristics of individual trading environments.

It is essential to recognise the limits of these models. Some solutions [33–35] solely deploy anonymisation techniques to provide privacy risk, as it allows for inferences about behaviours and schedules. The information, including how much energy is traded and at what price, can lead to sensitive insights about consumer behaviour, raising concerns about the protection of private information [8,19–22].

Privacy of P2P market participants may be compromised if their sensitive trading data is not protected, risking unauthorised leaks or sharing, potentially breaching GDPR standards [23]. Hence, ensuring secure and privacy-preserving data processing and exchange is privacy. However, it is crucial to emphasise that these techniques can be vulnerable to reverse engineering attacks, potentially leading to the exposure of private information by using techniques described in [50]. Some systems, as described in [36–39], are vulnerable to a Single Point of Failure (SPF), which can result in system-wide failures and outages. Moreover, as indicated in [40,41], some systems lack scalability, limiting their capacity to manage many transactions or users. Furthermore, it is worth noting that certain models are not fully optimised, as indicated in [42]. This could lead to sub-optimal performance or inefficiencies in their operation. Similarly, some solutions do not provide a comprehensive evaluation of their performance, as stated in [51]. Some of the other works [44–48] deployed Differential Privacy (DP) as a privacy-enhancing technique in their studies. However, in DP, data accuracy is sacrificed in exchange for improved privacy [15]. Among the privacy-preserving game theoretical energy trading solutions which use Homomorphic Encryption (HE) as a privacy-enhancing technique, Xia et al. [49] proposes a nested market clearing algorithm using distributionally robust optimisation and homomorphic encryption to secure privacy in energy trading. However, potential privacy breaches during inter-microgrid trading are not entirely eliminated due to reliance on anonymisation and desensitisation. The solution [25] provides full data privacy without compromising data accuracy, and SPF is avoided with decentralisation. However, the clearance mechanism provided in this work is not competitive, which is not a realistic scenario and does not incentivise users. The clearance mechanism in our previous work [8] is competitive, which is an incentive for the users, but its privacy-preserving framework falls short of complete optimisation in terms of computational efficiency. This limitation becomes apparent when handling a constrained number of participants within the trading platform.

In summary, the current state-of-the-art work reveals limitations across privacy, security, computational efficiency, and scalability aspects, especially within clearance mechanisms in LEMs. Unfortunately, there is no single solution that adequately tackles all these challenges simultaneously.

To address these limitations in the literature, we propose a computationally efficient Privacy-Preserving clearance mechanism for Local Energy Markets (PP-LEM). Our approach not only overcomes the inherent computational inefficiencies observed in existing frameworks but also enhances users' privacy. Crucially, PP-LEM outperforms the computational performance of previous solutions like [8], all while ensuring users' social welfare remains intact throughout the trading process.

The main motivation of this paper is to address the primary barriers hindering the broad adoption of LEMs, where entities utilise their own RES. These barriers include privacy risks, computational inefficiencies, and insufficient economic incentives for participants. By addressing these challenges, we aim to develop a privacy-preserving, computationally efficient, and economically efficient framework for energy trading in LEMs. This approach will promote a sustainable and user-centric integration of RES into the energy grid by incentivising users with RES to participate in the system through the benefits offered by our PP-LEM. Ultimately, this will reduce dependence on fossil fuels and enhance energy resilience.

This work introduces the following novel contributions:

- We have designed and implemented a novel *competitive* game theoretical clearance mechanism, modelled as the Stackelberg Game. The clearance mechanism is designed so that computationally efficient privacy-enhancing mechanisms can be employed.
- Building upon the devised clearance mechanism, we have developed and implemented a privacy-preserving clearance mechanism for LEMs. This mechanism utilises a computationally efficient cryptographic system, specifically incorporating a lightweight combination of partially homomorphic encryption and specialised encrypted square operations. These measures are aimed at protecting the sensitive information of both buyers and sellers in a computationally efficient way.

- We have performed an extensive evaluation of the proposed systems encompassing privacy, the existence of a Nash equilibrium, computational and communication cost analysis, and comprehensive experimental simulation results, which include data generation. The simulations involve deploying battery storage and running various configurations encompassing different dates.

Paper organisation: Background and Related Work is presented in Section 2. Design Preliminaries, including system model & iterations, threat model & assumptions, and requirements, are given in Section 3. Section 4 presents the proposed *Computationally Efficient Privacy Preserving Clearance Mechanism for Local Energy Trading Markets* (PP-LEM). Section 5 provides a comprehensive evaluation of PP-LEM through both theoretical analysis and experimental testing. In Section 6, we compare PP-LEM with existing privacy-preserving clearance mechanisms. Practical applications and limitations of PP-LEM are discussed in Section 7. Finally, Section 8 concludes the paper and gives directions for future work.

## 2. Background and related work

### 2.1. Background

Within LEMs, pricing and clearance mechanisms mainly rely on Auction Theory and Game Theory [29]. Auction theory, a branch of economics, focuses on auction design and analysis to understand bidder and seller behaviour. One key concept in auction theory is the double auction, where buyers and sellers submit prices they are willing to buy or sell. Transactions occur when bid and ask prices match [52]. Another important mechanism is the Vickrey–Clarke–Groves (VCG) mechanism, a type of auction where bidders submit true valuations. The winner is determined based on these valuations, and payments are designed to ensure truthfulness [53]. On the other hand, game theory is a mathematical framework used to study strategic interactions between rational decision-makers [54]. In the context of game theory, the Stackelberg game is a model where one player, the leader, makes the first move, and the other player, the follower, observes this move before making their own decision [54]. This asymmetry in information can lead to different outcomes compared to simultaneous-move games. Additionally, the Nash game, based on the concept of Nash Equilibrium, involves players making decisions while taking into account the decisions of others, with no player having an incentive to unilaterally change their strategy [54]. Auction and game theory inform solutions for determining prices in the clearance phase of LEMs.

In the context of privacy considerations within LEMs, protecting user privacy involves employing several key techniques. These techniques include anonymisation, Multi-Party Computation (MPC), Differential Privacy (DP), and Homomorphic Encryption (HE). Anonymisation entails the modification or removal of personally identifiable information from datasets to prevent the identification of individuals [55]. MPC enables multiple parties to collectively compute a function over their inputs while ensuring the confidentiality of those inputs [56]. DP involves the intentional modification or perturbation of data to obscure individual information while retaining the ability to manipulate data within a specified scope [57]. Lastly, HE allows computation on encrypted data without the need for prior decryption, thereby preserving privacy throughout the computation process [58]. These privacy-enhancing techniques play a crucial role in addressing the privacy concerns associated within LEMs.

### 2.2. Related work

Existing solutions which explored privacy-preserving market clearance mechanisms for LEMs [8,25,33–38,40–42,44,46–48,51] are summarised in Table 1 with their trading methods and privacy-preserving techniques used.

**Table 1**
Privacy Preserving Clearance Mechanisms with Their Privacy Enhancing Techniques.

| Paper | Approach | Clearance methods | Privacy enhancing techniques |
|---|---|---|---|
| Zhang et al. *2017* [33] | Auction Theory | Double Auction | Anonymisation |
| Li et al. *2018* [34] | Auction Theory | Double Auction | Anonymisation |
| Li et al. *2018* [35] | Auction Theory | Double Auction | Anonymisation |
| Sarenche et al. *2021* [36] | Auction Theory | Double Auction | HE & Anonymisation |
| Li et al. *2017* [37] | Auction Theory | Double Auction | HE |
| Yang et al. *2020* [38] | Auction Theory | Double Auction | HE |
| Bevin et al.*2023* [39] | Auction Theory | Double Auction | HE |
| Abidin et al. *2016* [40] | Auction Theory | Double Auction | MPC |
| Zobiri at al. *2022* [41] | Auction Theory | Double Auction | MPC |
| Zobiri at al. *2022* [42] | Auction Theory | Double Auction | MPC |
| Wang at al. *2023* [43] | Auction Theory | Double Auction | HE & MPC |
| Li et al. *2019* [44] | Auction Theory | Double Auction | DP |
| Yu et al. *2024* [45] | Auction Theory | Double Auction | DP |
| Hassan et al. *2019* [46] | Auction Theory | Vickrey–Clarke–Groves | DP |
| Hoseinpour et al. *2023* [47] | Auction Theory | Vickrey–Clarke–Groves | DP |
| Li et al. *2023* [48] | Game Theory | Stackelberg Game | DP |
| Xia et al. *2022* [49] | Game Theory | Nash Game | HE & Anonymisation |
| Xie et al. *2020* [25] | Game Theory | Stackelberg Game | HE |
| Erdayandi et al. *2022* [8] | Game Theory | Stackelberg Game | HE |

HE: Homomorphic Encryption, MPC: Multi-Party Computation, DP: Differential Privacy.

A privacy-preserving double auction mechanisms have been introduced by Zhang et al. [33] and Li et al. [34], specifically designed for scenarios wherein the requested and consumed power do not match, necessitating the exchange of surplus energy among households. The double auction mechanism is preceded by the households' acquisition of tokens from energy providers designated for subsequent trading activities. Privacy enhancement involves an anonymisation technique with data fragmentation into randomised segments, each encrypted with distinct key pairs to shield data owner identities during token reuse. Market operations are performed by a centralised control centre, adapting to user needs. However, the centralisation introduces a potential SPF. Moreover, the employment of anonymisation techniques carries inherent risks, as there is a possibility of the reverse engineering of private data [50].

Li et al. [35] propose a Demand Response (DR) mechanism for microgrids utilising vehicle-to-vehicle technology within the smart grid, emphasising location privacy protection through the Internet of Vehicles. Employing an online double auction, the mechanism supports efficient energy trading among electric vehicles (EVs) with surplus or deficit energy, accounting for each participant's utility. A k-anonymity anonymisation technique has been implemented to protect sensitive location data. Nonetheless, as mentioned earlier, it is imperative to recognise the potential privacy risks associated with anonymisation techniques.

In ensuring the secure and privacy-preserving implementation of diverse double auction mechanisms within smart grids, Sarenche et al. [36] introduce security protocols. Each participant is endowed with a pseudo-identity to guarantee anonymity, coupled with the encryption of bids utilising the Paillier cryptosystem — a HE technique strategically chosen to protect user privacy. The verification of participants' adherence to the auction protocol is conducted by utilising the Pedersen commitment scheme. However, the envisaged framework exhibits limitations concerning its applicability to decentralised applications, given that a central entity performs the majority of operations. Consequently, this centralised structure introduces susceptibility to a potential SPF.

In addressing the issue of DR for EVs during microgrid outages, the works by Li et al. [37] and Yang et al. [38] investigate the application of Vehicle-to-Grid (V2G) technology. Notably, they propose a privacy-preserving double auction scheme. The MicroGrid Centre Controller (MGCC) assumes the role of an auctioneer in the market, efficiently solving the social welfare maximisation problem by matching buyers to sellers. To protect privacy, a cloud intermediary employs HE to facilitate communication between bidders and the auctioneer. Furthermore, Yang et al. [38] introduce an additional layer of protection for sensitive location information. Despite these advancements, both Li et al. [37], and Yang et al. [38] encounter a potential drawback: all double auction operations are centralised within the MGCC, posing a risk of an SPF in these mechanisms. Moreover, Bevin et al. [39] proposes a privacy-preserving double auction model for local electricity markets using the Paillier HE system. Cluster Coordinators aggregate household data in an encrypted format, which is then transmitted to the Market Operator. The Market Operator decrypts the aggregated data and performs the double auction in plaintext format. However, since the Market Operator has access to the plaintext data of clusters, there is a risk of inferring household data from these clusters. Additionally, because the main double auction operations are conducted by the Market Operator, there is a risk of SPF.

Within the realm of secure and privacy-preserving mechanisms for electricity trading, Abidin et al. [40] introduce a bidding system founded on secure MPC. This approach allows mutually distrustful parties to engage in computations without revealing sensitive data. The trading platform orchestrates a double auction to ascertain trade parameters such as price, volume, and auction winners. However, the performance of the system could be further enhanced, particularly in the context of short trading periods. Furthermore, the scalability of the system in relation to communication overhead remains to be evaluated. Similarly, Zobiri et al. [41] present a privacy-preserving demand response market with an incentive structure for users trading their flexibility. User offers are aggregated over encrypted data using MPC to uphold user privacy. Double auctions govern the distribution of flexibility and consumer consumption schedules, though the computational cost warrants further examination. The offline phase's performance is unspecified, and online phase computations require optimisation for trading periods shorter than 30 minutes. Likewise, Zobiri et al. [42] propose a privacy-preserving energy trading market where users exchange excess electricity and flexibility. Offers/bids are submitted in encrypted form, and MPC is employed for computations, considering device constraints for resource allocation. However, the cost of communication still needs to be explored.

Moreover, Wang et al. [43] explores a new trading strategy that enhances the privacy of participants in blockchain-based P2P electricity markets. The authors propose a three-layer architecture composed of a physical layer, an information layer, and a market layer to protect sensitive trading data. This strategy utilises HE and MPC to secure on-chain data, enabling safe and confidential market operations. Despite utilising HE, a computationally intensive privacy-enhancing technique, the system's performance has only been examined for a low number of users (e.g., 50 users), and detailed performance results have not been reported. Similarly, while employing MPC, which is communication-intensive, the communication costs have not been investigated.

To address DR in smart grids with island MicroGrids (MGs), Li et al. [44] propose a differential-privacy-based auction market. EVs with surplus energy act as sellers, while those with insufficient energy act as buyers in a double auction system. The MGCC functions as the auctioneer, performing computations with differential privacy for added privacy. Relying on the MGCC for computations introduces a potential risk of SPF, and the deployment of differential privacy, while enhancing privacy, comes at the cost of some degree of accuracy.

The work Yu et al. [45] introduces an approach to managing virtual power plants (VPPs) by leveraging blockchain technology. It proposes a proof of comprehensive energy ratio (PoCER) consensus mechanism to enhance the participation of renewable energy sources in VPP auctions, coupled with a sharding consensus strategy to improve scalability and efficiency. The study integrates Rényi differential privacy and Dirichlet differential privacy methods to protect sensitive data during auction and consensus processes, ensuring both security and efficiency. However, implementing differential privacy improves privacy but can slightly compromise accuracy.

In the work by Hassan et al. [46], a private auction model is proposed for energy trading in blockchain-based microgrids. The use of Blockchain technology successfully mitigates SPF risks. The VCG mechanism is adapted for auction establishment, incorporating differential privacy for enhanced privacy protection. While the modified VCG mechanism proves superior in generating revenues and overall network benefits compared to the original version, the adoption of differential privacy unavoidably introduces a trade-off with data accuracy.

In Hoseinpour et al. [47], a differentially private VCG mechanism is introduced for local energy markets. While this approach enhances privacy through differential privacy, there is a trade-off as some degree of social welfare is sacrificed due to the introduced perturbations. Similarly, Li et al. [48] presents a privacy-preserving two-stage single-leader multi-follower Stackelberg game for data markets, leveraging DP for enhanced privacy protection. Nevertheless, this privacy-enhancing strategy comes at the cost of reduced accuracy in the data, introducing a trade-off between privacy and accuracy.

Xia et al. [49] proposes a nested market clearing algorithm using distributionally robust optimisation and homomorphic encryption to secure privacy in energy trading. Privacy is preserved by encrypting sensitive parameters of prosumers during the ADMM process, preventing the exposure of sensitive information while allowing efficient optimisation. However, potential privacy breaches are not entirely eliminated due to reliance on anonymisation and desensitisation, and the efficiency and accuracy of these algorithms depends on proper parameter tuning, which can be challenging in real-world applications.

In work by Xie et al. [25], a framework for preserving privacy in distributed energy trading is presented, employing a game theoretical approach. The approach involves buyers privately computing optimal trading prices and sellers allocating trading amounts through HE without revealing sensitive data. The trading scenario is formulated as a non-cooperative Stackelberg game, where buyers act as leaders to determine the optimal price, followed by sellers as followers to determine trading amounts. The decentralised nature of the computations eliminates the dependence on a centralised entity, mitigating the risk of an SPF. However, the designed market lacks competitiveness, as a fixed market price is determined by buyers, and trading occurs at this predetermined price, creating an unrealistic scenario that fails to incentivise users.

To overcome this limitation, we proposed in our previous work [8] a Privacy-Friendly *competitive* Energy Trading platform (PFET) based on game theory, specifically Stackelberg competition, which incentivises users. PFET introduces a *competitive* market where prices and demands are established through competition and computations are carried out in a decentralised manner, free from reliance on a central authority. Although simulations demonstrate the efficacy of PFET for local communities with up to 100 users, scalability remains a challenge.

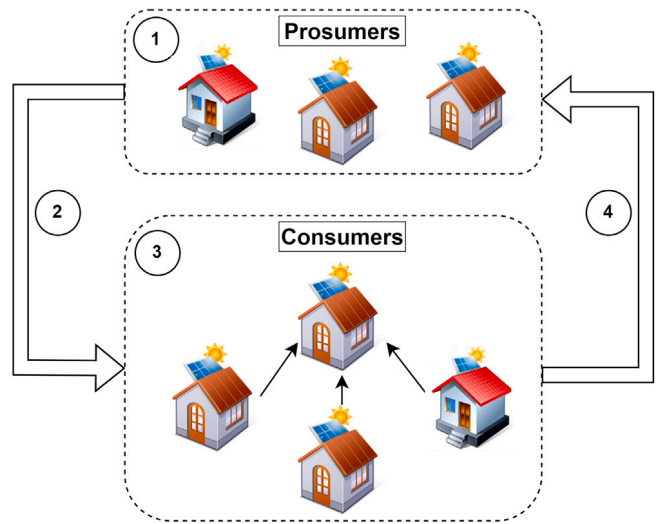

**Fig. 1.** System model and iterations.
**Interactions in circles:** 1) Proposal of prices by prosumers. 2) Transfer of prices from prosumers to consumers. 3) Calculation of strategies by consumers. 4) Transfer of reactions of consumers to prosumers.

Addressing this scalability issue, our approach focuses on improving the algorithm's computational efficiency instead of resorting to additional hardware or embedded systems, as suggested by alternative methods [59–62]. Our proposed Privacy-Preserving clearance market for Local Energy Markets (PP-LEM), detailed in Section 4, surpasses the computational inefficiencies of the prior framework [8] while providing privacy for the users. Significantly, PP-LEM not only outperforms the computational efficiency of [8] but also ensures that users' social welfare remains uncompromised throughout the trading process.

## 3. Preliminaries

### 3.1. System model and iterations

Our proposed clearance mechanism for LEMs is depicted in Fig. 1. This mechanism utilises a Peer-to-Peer (P2P) energy trading model, facilitating decentralised energy transactions. In the P2P model, energy producers, typically households with renewable energy sources like solar panels, can directly sell excess energy to consumers [10].

The system operates with two main player types: prosumers and consumers. Prosumers equipped with RES, such as solar panels, not only fulfil their energy needs with RES but also contribute surplus energy to consumers. Conversely, consumers, also equipped with RES, resort to purchasing electricity when they cannot meet their energy requirements through self-generation with RES. Each household is equipped with a Smart Meter (SM) to measure imported and exported electricity [28,63]. Prosumers are designated as 'sellers' when they have excess electricity after meeting their own energy demands, while consumers are classified as 'buyers' when they cannot fulfil their energy needs internally. Only participants capable of selling and buying, respectively, are involved in the P2P clearance mechanism.

The primary aim of the clearance mechanism is to establish final prices at which prosumers can sell specific amounts of energy and to align supplies with demands. This mechanism employs a Stackelberg game, where buyers and sellers form leader and follower teams, respectively. Initially, sellers propose strategies by adjusting their proposals in response. As the market progresses towards equilibrium, sellers adapt to buyer reactions, and buyers align their demands accordingly.

The interactions in our model, denoted as ①, ②, ③, and ④ in Fig. 1, are elaborated as follows: Sellers initiate the process by offering selling prices for excess electricity in the first iteration ①, transmitting these proposals to buyers through a communication channel ②. Each buyer formulates individual strategies, which are aggregated by a buyer representative ③. Aggregated strategies as reactions to each seller are then communicated back ④. Iterations persist until a market equilibrium point is attained, where supply aligns with demand. At market equilibrium, the final prices at which prosumers can sell specific amounts of energy are established.

### 3.2. Threat model and assumptions

In this subsection, threat model and assumptions are presented.

#### 3.2.1. Threat model

*Consumer entities*: They are assumed to be honest but curious about prosumers' sensitive data. Adhering to protocol specifications, they may still attempt to acquire the confidential information of individual prosumers.

*Prosumer entities*: They are also assumed to be honest but curious for consumers' sensitive data.

*External entities*: External entities are not trustworthy. They might engage in eavesdropping on data during transit or intercept and manipulate the transmitted information.

#### 3.2.2. Assumptions

*Communication Channel:* We make the assumption that the entities engage in communication through channels that are both secure and authentic [63].

*Smart Meters:* They are produced to be invulnerable to tampering and securely sealed; these entities ensure that any attempt at manipulation, even by their users, would be promptly detected [64].

*Certificates:* A trusted authority certifies the public keys. Each user has knowledge of the certificates belonging to other users [63].

*Rationality:* By participating in PP-LEM, users aim to increase their benefits. Consumers focus on lowering their costs and increasing their utility, while sellers aim to increase their total profit [63].

*Trading Period:* Financial reconciliation takes place during settlement cycles (SCs) for users who participated in trading [28]. Within a SC, the prices and quantities for buying or selling electricity are established for a designated time period. One SC is assumed to be one hour.

### 3.3. Requirements

In this section, functional and privacy requirements are presented.

#### 3.3.1. Functional

*Social welfare provision*: The primary goal of PP-LEM should lie in optimising social welfare, seeking to maximise the utilities of users to the greatest extent possible.

*Individual rationality*: PP-LEM should maintain individual rationality, ensuring that agents receive higher payoffs for participating than they would for abstaining.

*Market equilibrium*: PP-LEM should reach to a market equilibrium, where supply aligns with demand.

#### 3.3.2. Privacy

PP-LEM must refrain from disclosing any personal user information.

*Seller price confidentiality*: Seller pricing information must remain confidential from buyers, as the disclosure of prices can unveil sensitive data regarding the electricity consumption or production patterns of the sellers. As an example, the selling price of excess electricity of the seller prosumers will be high when they consume a high amount of energy.

*Demand confidentiality*: The aggregate demand of buyers on a specific seller needs to be computed in a manner that preserves privacy, allowing buyers to perform reaction calculations without disclosing individual demand data on a specific seller. The revelation of demand information can expose sensitive details about the production or consumption patterns of prosumers.

*Buyer profile variables confidentiality*: Buyer-specific sensitive profile variables that reveal information about electricity consumption/generation patterns should be hidden from sellers.

### 3.4. Building blocks

In this subsection, we introduce the foundational components essential for a privacy-preserving energy trading platform. Our system relies on HE to uphold privacy standards. HE stands out as the optimal choice, ensuring data accuracy with minimal communication intensity while maintaining data integrity. Specifically, we have implemented Partially Homomorphic Encryption (PHE), a computationally lightweight variant of HE utilising the Paillier Cryptosystem [65], as described in Section 3.4.1. Although its computational overhead is lower compared to Fully Homomorphic Encryption (FHE), PHE supports a restricted set of operations — limited to addition and scalar multiplication. To overcome the limitations of the Paillier Cryptosystem, which only supports addition and scalar multiplication, we have integrated a specialised Squaring Operation scheme for encrypted data computations, as briefly described in Section 3.4.2.

#### 3.4.1. Paillier cryptosystem

The Paillier Cryptosystem [65] comprises of three fundamental functions for key generation ($KGen_{pe}$), encryption ($Enc_{pe}$), and decryption ($Dec_{pe}$).

- $KGen_{pe}(k) \rightarrow PK, SK$: The key generation function takes a security parameter $k$ as input and generates a public–private key pair $PK$ and $SK$. The function selects two large prime numbers $p$ and $q$ given the security parameter $k$, computes $n = p \cdot q$, and $\lambda = lcm(p - 1, q - 1)$. It selects a generator $g \in \mathbb{Z}^*_{n^2}$ where $\mathbb{Z}^*_{n^2}$ denotes the Set of integers relatively prime to $n^2$, and computes $\mu = (L(g^\lambda \mod n^2))^{-1} \mod n$ where $L(x) = (x-1)/n$. The resulting public key is $PK = (n, g)$ and the private (secret) key is $SK = (\lambda, \mu)$.
- $Enc_{pe}(PK, m) \rightarrow c$: The encryption function takes a plain message $m \in \mathbb{Z}$ and a public key $PK = (n, g)$ as input and generates a ciphertext $c$. It selects a random number $r \in \mathbb{Z}^*_n$, where $\mathbb{Z}^*_n$ denotes the set of integers relatively prime to $n$ and computes $c = g^m \cdot r^n \mod n^2$.
- $Dec_{pe}(SK, c) \rightarrow m$: The decryption algorithm takes a ciphertext $c$ and a private key $SK = (\lambda, \mu)$ as input and outputs the message $m$. It recovers the message as $m = L(c^\lambda \mod n^2) \cdot \mu \mod n$.

The Paillier Cryptosystem permits two main arithmetic operations over encrypted data. These operations are Addition ($Add_{pe}$) and Scalar Multiplication ($Smul_{pe}$) respectively.

---

**Algorithm 1:** Square

**Input:** $\{X\}_{\mathcal{E}} = Enc_{pe}(PK, X)$
**Output:** $\alpha, \omega$

1   $r \leftarrow Rand()$;
2   $tmp_1 \leftarrow \{-r\}_{\mathcal{E}} = Enc_{pe}(PK, -r)$;
3   $\omega \leftarrow \{X - r\}_{\mathcal{E}} = Add_{pe}(\{X\}_{\mathcal{E}}, tmp_1)$;
4   $tmp_2 \leftarrow \{-r^2\}_{\mathcal{E}} = Enc_{pe}(PK, -r^2)$;
5   $tmp_3 \leftarrow \{2rX\}_{\mathcal{E}} = Smul_{pe}(2r, \{X\}_{\mathcal{E}})$;
6   $\alpha \leftarrow \{-r^2 + 2rX\}_{\mathcal{E}} = Add_{pe}(tmp_2, tmp_3)$;

---

### 3.4.2. Squaring operation

The Square operation over encrypted data is outlined in Algorithm 1. The procedure involves squaring an encrypted version of variable $X$ denoted as $\{X\}_{\mathcal{E}} = \mathsf{Enc}_{pe}(PK, X)$. In this work, the notation $\{.\}_{\mathcal{E}}$ refers to a homomorphically encrypted data. Here is a step-by-step explanation: A random number $r$ is selected in line 1, and its negative is encrypted in line 2. This encrypted negative value is added to the Encrypted $X$ denoted as $\{X\}_{\mathcal{E}}$, in line 3, resulting in an encrypted version of $X - r$ presented as $\omega \leftarrow \{X - r\}_{\mathcal{E}}$. The negative of $r^2$ is encrypted in line 4, and $\{X\}_{\mathcal{E}}$ is multiplied by $2r$ in line 5. Finally, the results from lines 4 and 5 are added together to obtain the encrypted version of $-r^2 + 2rX$ presented as $\alpha \leftarrow \{-r^2 + 2rX\}_{\mathcal{E}}$ in line 6. The Squaring operation outputs $\alpha$ and $\omega$, both encrypted parameters. These encrypted parameters, $\alpha$ and $\omega$, can be decrypted and processed to obtain the plaintext $X$ again. The correctness of the Squaring operation is proven by Theorem 1 below.

**Theorem 1.** *Squaring operation is correct.*

$$\mathsf{Enc}^{-1}(Square(\{X\}_{\mathcal{E}})) = X^2 \tag{1}$$

*where*

$$\mathsf{Enc}^{-1}(\alpha, \omega) = \mathsf{Dec}_{pe}(PK, \alpha) + \mathsf{Dec}_{pe}(PK, \omega)^2 \tag{2}$$

**Proof.**

$$\mathsf{Dec}_{pe}(SK, \alpha) = -r^2 + 2rX \tag{3}$$

$$\mathsf{Dec}_{pe}(SK, \omega)^2 = (X - r)^2 = X^2 - 2rX + r^2 \tag{4}$$

Adding Eqs. (3) and (4) results in

$$-r^2 + 2rX + X^2 - 2rX + r^2 = X^2 \tag{5}$$

## 4. Computationally efficient privacy-preserving clearance mechanism for local energy markets (PP-LEM)

In this section, we introduce our novel Privacy Preserving Clearance Mechanism for Local Energy Markets (PP-LEM). Firstly, we present our novel *competitive* Game Theoretical Clearance Mechanism, modelled as a Stackelberg Game. This clearance mechanism is meticulously designed to facilitate the integration of computationally efficient privacy-enhancing mechanisms. Secondly, we expand with the proposed clearance mechanism with privacy enhancing technologies — leveraging the devised clearance mechanism, we have developed and implemented a Privacy-Preserving Clearance Mechanism with Homomorphic Encryption. This model harnesses a computationally efficient cryptosystem, specifically incorporating a lightweight combination of partially homomorphic encryption and specialised encrypted square operations.

### 4.1. Game theoretical clearance mechanism

Our proposed clearance mechanism leverages Game Theory, specifically the Stackelberg Game, to establish the selling prices of electricity by sellers while aligning demand and supply in the P2P market. As illustrated in Fig. 1, the mechanism employs an iterative model. Initially, a seller — the prosumer with excess electricity, proposes an initial price offer. Then, the buyer —, the consumer in need of electricity, responds to this offer. Subsequent iterations involve sellers adjusting their price proposals based on the buyers' reactions, with the process continuing until the supply and demand equilibrium is achieved.

The mechanism is operationalised through two distinct algorithms: the Seller's Algorithm (see Algorithm 2) and the Buyer's Algorithm (refer to Algorithm 3). In this Stackelberg Game framework, sellers aim to increase their revenue through non-cooperative and competitive

---

**Algorithm 2:** Sellers' Algorithm

**Input:** Number of Sellers, Buyers $[N_S, N_B]$

1   initialisation;
2   **for** *Time* $t_k$ **do**
3     $D_{Tot} \leftarrow receiveTotalDemand();$
4     $S_{Tot} \leftarrow calculateTotalSupply();$
5     **for** *Each* $s_j$ **do**
6       **if** $S_{Tot} > D_{Tot}$ **then**
7         $S_{Tot}^{P2P} \leftarrow D_{Tot};$
8         $S_j^{P2P} \leftarrow \frac{(S_j \times D_{Tot})}{S_{Tot}};$
9       **else**
10        $S_{Tot}^{P2P} \leftarrow S_{Tot};$
11        **for** *Each* $s_j$ **do**
12         $S_j^{P2P} \leftarrow S_j;$
13        **end**
14       **end**
15     **end**
16     $Prices \leftarrow ProposePrices[\pi_1, \pi_2, ..., \pi_{N_S}];$
17     **do**
18       $[W_{B_1}, ..., W_{B_{N_S}}, W_{Tot}] \leftarrow Buyers'Algorithm(Prices);$
19       **for** *Each* $s_j$ **do**
20        $D_j \leftarrow \frac{S_{Tot}^{P2P} \times W_{B_j}}{W_{Tot}};$
21        $\pi_j \leftarrow \pi_j + \eta_1 \times (D_j - S_j^{P2P}) ;$
22        $\pi_j \leftarrow min(\rho_{Sup}, max(\rho_{FiT}, \pi_j))$
23       **end**
24     **while** $|D_j - S_j^{P2P}| > \epsilon$, *For any* $s_j$;
25   **end**

---

**Algorithm 3:** Buyers' Algorithm

**Input** : Prices $[\pi_1, \pi_2, \pi_3, ..., \pi_{N_S}]$
**Output:** Welfares $[W_{B_1}, W_{B_2}, W_{B_3}, ..., W_{B_{N_S}}, W_{Tot}]$

1   **for** *Each* $s_j$ **do**
2     **for** *Each* $b_i$ **do**
3       $X_{ji} \leftarrow (\lambda_i - \pi_j)/\theta_i;$
4     **end**
5     $W_{B_j} \leftarrow \frac{1}{2} \times \sum^i \theta_i \times X_{ji}^2 ;$
6   **end**
7   $W_{Tot} \leftarrow \sum^J W_{B_J} ;$

---

strategies as delineated in Algorithm 2. Conversely, buyers strive to optimise their utility by securing the desired amount of energy at the best possible terms, as facilitated by Algorithm 3.

This mechanism is initially presented without considering privacy concerns in this part, and subsequently, a privacy-preserving variant is detailed in Section 4.2. It is essential to note that this non-privacy-privacy version of the mechanism in this part is designed for computational efficiency for the privacy-preserving version to allow incorporating a lightweight partially homomorphic cryptosystem and special homomorphic square operation. This design choice ensures that once data is encrypted, only scalar multiplication and squaring operations are performed on the encrypted data, thereby avoiding the computational burden of multiple encrypted data multiplications with fully homomorphic encryption.

The system initialisation is primarily done by the sellers, utilising Algorithm 2. The buyers' algorithm (see Algorithm 3) is invoked within the framework of the sellers' algorithm, ensuring a cohesive interaction between the two parties. The following sections detail the Sellers' Algorithm and the Buyers' Algorithm, respectively, providing a comprehensive understanding of their roles and functionalities within the system.

### 4.1.1. Sellers' algorithm

In the Sellers' algorithm, a concise series of steps are executed to facilitate the clearance mechanism in P2P setting: *(1) Calculation of Supplies (Lines 3–15):* Initially, the algorithm calculates the supplies each seller can offer to the P2P market and total supply of sellers denoted as $S_j^{P2P}$ for each seller $s_j$ and $S_{Tot}^{P2P}$ respectively. *(2) Initial Pricing (Line 16):* Sellers propose initial prices for their supplies. *(3) Price Feedback (Line 18):* These initial prices are sent to the Buyers' algorithm. Buyers then evaluate these prices, calculate their reaction, and send feedback to the sellers. *(4) Price Adjustment and Market Equilibrium (Lines 19–24):* Sellers adjust their prices based on buyer feedback, iteratively refining them until market demands align with supplies, signifying market equilibrium at line 24.

Now, each step is explained in detail:

*(1) Calculation of Supplies:* In the context where sellers are prosumers with surplus electricity – defined as the electricity generated minus the energy they consume – the process starts with an initialisation step. After this, for each trading period, two critical figures are established: Total Demand ($D_{Tot}$), which is the total energy buyers wish to purchase is received from buyers, and Total Supply ($S_{Tot}$) is calculated by summing the surplus electricity from all sellers.

The core aim of the algorithm is to ensure an effective match between supply and demand by the end of each trading period. Early in the process, between lines 5 and 15, the algorithm determines the amount of electricity each seller can trade in the P2P market, denoted as $S_j^{P2P}$, along with the total energy available for trading, $S_{Tot}^{P2P}$.

When $S_{Tot}$ exceeds $D_{Tot}$, indicating sellers have more surplus electricity than buyers wish to purchase, a mismatch occurs. Sellers are then limited to selling only up to the total demand. Consequently, $S_{Tot}^{P2P}$ is set to $D_{Tot}$ and the individual seller's available surplus for sale, $S_j^{P2P}$, is adjusted to reflect this, with each unit of surplus energy being proportionally rationed based on the ratio of $D_{Tot}/S_{Tot}$. This ensures that the distribution of electricity to be sold is aligned with the overall market demand, preventing any seller from expecting to sell more than what the market can absorb.

Conversely, if $S_{Tot}$ is less than $D_{Tot}$, the total demand is adjusted to match the total supply, recognising that buyers' demand cannot exceed what sellers can provide. Here, $S_{Tot}^{P2P}$ is set to $S_{Tot}$, and each seller's surplus, $S_j$, is then allocated for trading as $S_j^{P2P}$, ensuring that the market operates efficiently without exceeding its supply limits.

*(2) Initial Pricing (Line 16):* After obtaining $S_{Tot}^{P2P}$ and $S_j^{P2P}$ for each seller, all sellers propose their initial prices $[\pi_1, \ldots, \pi_{N_S}]$ to sell their excess energies in line 16.

After this, the competition – iterations start to determine the final price of each user. Throughout the process, starting from the initial iteration to the final iteration, prices reach their market equilibrium points where Demands and Supplies match with each other.

*(3) Price Feedback (Line 18):* In this non-cooperative Stackelberg Game, after sellers propose prices in line 16, they are then transmitted to buyers in line 18. Subsequently, buyers compute reaction functions using the Buyer's algorithm (Algorithm 3) in response to the offered prices. Reactions functions of buyers inside the Buyer's algorithm outputs Welfares $[W_{B_1}, W_{B_2}, W_{B_3}, \ldots, W_{B_{N_S}}]$ – payoffs of all buyers obtained from each seller $s_j$ for $j = 1$ to $N_S$. The welfares derived from purchasing a specific amount of energy at a particular price from a given seller are correlated to the demand placed on that specific seller. The more benefits (welfare) buyers receive from a particular seller, the more electricity they demand from the seller. Total Welfare $W_{Tot}$ – welfare obtained from all sellers is also calculated inside the Buyer's algorithm. Then, calculated welfares and $W_{Tot}$ are returned to sellers from Buyer's algorithm.

*(4) Price Adjustment and Market Equilibrium (Lines 19–24):* After receiving the welfares and $W_{Tot}$ from buyers in line 18 of Algorithm

2, for each seller, demand – $D_j$ is calculated in line 20, and the next price for the next iteration is calculated in accordance with lines 21 and 22. The iterations continue until supplies match the demands where $|D_j - S_j^{P2P}|$ is smaller than a stopping criteria $\epsilon$ for any seller $s_j$.

In line 20, $D_j$ for each $s_j$ is obtained by directly proportioning $S_{Tot}^{P2P}$ with $W_{B_j}/W_{Tot}$ ratio. In line 21, sellers decide on new prices for the next round based on the difference between what buyers demand from $s_j$ – denoted as $D_j$ and the maximum amount of power $s_j$ can provide in the P2P market, which is denoted as $S_j^{P2P}$. If buyers demand more than what is available, sellers raise the price; if there is more power than needed, sellers lower it. To prevent oscillations in prices and make the system reach a balanced state, they multiply this difference by a small number, $\eta_1$. On line 22, the updated prices are kept between the limits of $\rho_{FiT} < \pi_j < \rho_{Sup}$. The iterations continue until what buyers want aligns with what sellers can give, checked on line 24 for each $s_j$.

### 4.1.2. Buyers' algorithm

Within the Buyer's algorithm (Algorithm 3), the computation of buyer reactions which outputs expressed as Welfares $[W_{B_1}, W_{B_2}, W_{B_3}, \ldots, W_{B_{N_S}}]$ and $W_{Tot}$ is performed in response to the prices put forth by sellers. In the Buyers' algorithm, the tasks are outlined as follows: *(1) the calculation of $X_{ji}$,* which represents the amount of electricity buyer $b_i$ intends to purchase from seller $s_j$. *(2) the determination of Welfares* for each seller $s_j$ and the Total Welfare. These tasks are explained in detail below, with an emphasis on the foundational formulas integral to Algorithm 3.

*(1) The calculation of $X_{ji}$:*

The consumption utility function, expressed in Eq. (6), measures the satisfaction level of buyers after consuming a particular amount of energy.

$$u_i^c(x_n) = \lambda_i \times x_n - \frac{\theta_i}{2} \times x_n^2 \qquad (6)$$

Here, $\lambda_i$, related to the buyer's profile data, represents a prosumer preference parameter that characterises the behaviours of consumers. This parameter varies between different consumers and may also change over time. It represents the user's energy needs based on their behaviour. If a user requires a high amount of energy based on their behaviour, $\lambda_i$ will be higher, and vice versa. The $\theta_i$ is a constant parameter [66]. The $x_n$ presents the amount of energy consumed.

When a buyer, $b_i$, consumes a quantity of energy, $X_{ji}$, purchased at a price, $\pi_j$, from a seller, $s_j$, the net utility, $U_i$, is calculated by subtracting the cost of the energy ($\pi_j \times X_{ji}$) from the consumption utility function as follows in Eq. (7).

$$U_i = u_i^c(X_{ji}) - \pi_j \times X_{ji} \qquad (7)$$

Based on Eq. (7), the computation of the quantity of energy that a buyer, $b_i$, aims to purchase from $s_j$, denoted as $X_{ji}$, is initially calculated on the line 3 of the Buyer's algorithm (refer to Algorithm 3). The objective for buyers, $b_i$, is to maximise the net utility function, $U_i$, as outlined in (7) concerning the price $\pi_j$ offered by the seller, $s_j$. The determination of the optimal value of $X_{ji}$ that maximises $U_i$ involves calculating the derivative of $U_i$, setting it equal to zero, and solving for $X_{ji}$.

$$U_i' = (\lambda_i - \pi_j) - \theta_i \times X_{ji} = 0 \qquad (8)$$

Solving for $X_{ji}$ in Eq. (8) yields in the following equation,

$$X_{ji} = (\lambda_i - \pi_j)/\theta_i \qquad (9)$$

The Eq. (9) is used to calculate $X_{ji}$ maximising the $U_i$ in line 3 of Algorithm 3 after each buyer $b_i$, receive the price $\pi_j$ offered by seller, $s_j$.

The pricing strategy employed by $s_j$ exhibits an inverse relationship with $X_{ji}$ —— when prices are high, buyers tend to opt for a lesser quantity of energy, and conversely, when prices are low, buyers are inclined to purchase more. It is essential to note that the minimum value for $\lambda_i$ must be greater than the maximum retail price.

*(2) The determination of Welfares:*

Moving on, the Welfare function, denoted as (10), encapsulates the collective utilities accrued by all buyers as they procure electricity from $s_j$. The calculation of $W_{B_j}$ occurs in line 5 of Algorithm 3.

$$W_{B_j} = \sum^i U_i = \frac{1}{2} \sum^i \theta_i X_{ji}^2 \tag{10}$$

$W_{Tot}$ which is an aggregation of all welfares is calculated in line 7 of Buyer's algorithm. Within the Buyer's algorithm, while each $b_i$ independently calculates $X_{ji}$, aggregations for $W_{B_j}$ and $W_{Tot}$ are executed across the chosen buyers. Upon the completion of all computations within the Buyer's algorithm, the resulting Welfares $[W_{B_1}, \ldots, W_{B_{N_S}}]$ and $W_{Tot}$ are presented as outputs to the sellers in response to their pricing proposals.

### 4.2. Privacy-preserving clearance mechanism with homomorphic encryption

In this part, we present our privacy-preserving variation of the Game Theoretical Clearance Mechanism proposed in Section 4.1 – we employ efficient privacy enhancing techniques to the clearance mechanism proposed in Section 4.1, to protect the privacy of sellers and buyers in the clearance mechanism. Specifically, we utilise a combination of the computationally efficient Paillier Cryptosystem [65] and a specialised squaring algorithm, a variation of techniques from [67], which implement basic partially homomorphic encryption functions. This combination enables privacy-preserving computations that are also computationally efficient. Essentially, this mixture of techniques, rooted in the principles of HE, allows for the processing of encrypted data without exposing the actual data, ensuring privacy. When the encrypted results are decrypted, the outcome is consistent with what would have been achieved without any encryption, thereby maintaining both privacy and accuracy in the clearance mechanism.

Here, we present two algorithms designed to implement the clearance mechanism while preserving privacy: Sellers' Algorithm with Homomorphic Encryption (Algorithm 4) and Buyers' Algorithm with Homomorphic Encryption (Algorithm 5). In essence, these algorithms operate by allowing sellers to propose prices, encrypting them, and then transmitting them to buyers. Subsequently, buyers compute reaction functions over the encrypted data without having access to the plaintext information provided by the sellers.

Once buyers have computed their reaction functions over encrypted data, they send the results back to the sellers in encrypted format. Sellers then decrypt these outputs, calculate the next set of prices, and this process iterates continuously until supply and demand align.

For detailed explanations of these algorithms, please refer to Sections 4.3 and 4.4.

### 4.3. Sellers' algorithm with HE

In this Sellers' Algorithm with Homomorphic Encryption (Algorithm 4), following tasks are performed: *(1) Supply Calculations (Lines 3–15):* Initially, the algorithm calculates the supplies each seller can offer to the P2P market and total supply of sellers denoted as $S_j^{P2P}$ for seller $s_j$ and $S_{Tot}^{P2P}$ respectively. *(2) Initial Pricing (Line 16–17):* Sellers propose initial prices for their supplies and encrypt them. *(3) Price Feedback (Line 19):* These initial encrypted prices are sent to the Buyers' algorithm. Buyers then evaluate these encrypted prices, calculate their reaction over homomorphically encrypted data, and send feedback to the sellers in an encrypted format. *(4) Price Adjustment and Market Equilibrium (Lines 20–35):* Sellers decrypt the information where they received from the buyers, then they adjust their prices based on buyer feedback, iteratively refining them until market demands align with supplies, ensuring market equilibrium conditions at line 35.

*(1) Supply Calculations (Lines 3–15):* Prior to computing the supply of a seller ($S_j^{P2P}$) and the total supply ($S_{Tot}^{P2P}$) that can be offered to the P2P market, we first obtain the total demand ($D_{Tot}$) and total supply

---

**Algorithm 4:** Sellers' Algorithm with HE

**Input:** Number of Sellers, Buyers $[N_S, N_B]$

```
1  initialisation;
2  for Time t_k do
3      D_Tot ← receiveTotalDemand();
4      S_Tot ← calculateTotalSupply();
5      for Each s_j do
6          if S_Tot > D_Tot then
7              S_Tot^P2P ← D_Tot;
8              S_j^P2P ← (S_j×D_Tot)/S_Tot;
9          else
10             S_Tot^P2P ← S_Tot;
11             for Each s_j do
12                 S_j^P2P ← S_j;
13             end
14         end
15     end
16     Prices ← ProposePrices[π_1, π_2, ..., π_{N_S}];
17     {Prices}_ℰ ← Encrypt(PK_S, Prices);
18     do
19         [Alphas, Omegas] ← Buyers' Algorithm({Prices}_ℰ);
20         for Each s_j do
21             W_{B_j} ← Dec_pe(SK_S, A_{B_j});
22             for Each ω in Ω_{B_j} do
23                 W_{B_j} ← W_{B_j} + Dec_pe(SK_S, ω)^2;
24             end
25         end
26         W_Tot ← Dec_pe(SK_S, A_Tot);
27         for Each ω in Ω_Tot do
28             W_Tot ← W_Tot + Dec_pe(SK_S, ω)^2;
29         end
30         for Each s_j do
31             D_j ← (S_Tot^P2P × W_{B_j})/W_Tot;
32             π_j ← π_j + η_1 × (D_j − S_j^P2P);
33             π_j ← min(ρ_Sup, max(ρ_FiT, π_j))
34         end
35     while |D_j − S_j^P2P| > ϵ, For any s_j;
36  end
```

---

($S_{Tot}$) of excess electricity available. In the privacy-preserving process for calculating $D_{Tot}$ by buyers and $S_{Tot}$, $S_j^{P2P}$, and $S_{Tot}^{P2P}$ by sellers, a series of secure steps are executed. Initially, one user encrypts their data using their HE public key and transmits it to another user. Subsequently, the receiving user encrypts their data and homomorphically adds it to the received encrypted data. This augmented data is then forwarded and accumulated by other users iteratively until the entirety of the data is aggregated in an encrypted format. Once the aggregated data is obtained in encrypted form, the user who initially encrypted the first data proceeds to decrypt it using their private HE key. Through this meticulous process, the final values of $D_{Tot}$ and $S_{Tot}$ are derived by buyers and sellers respectively.

After computing $D_{Tot}$ and $S_{Tot}$ in a privacy-preserving manner, the final values of $D_{Tot}$ and $S_{Tot}$ become publicly accessible. Subsequently, each seller $s_j$ independently calculates $S_j^{P2P}$ and $S_{Tot}^{P2P}$ in plaintext format.

*(2) Initial Pricing (Line 16–17):* After the calculation of $D_{Tot}$ by buyers and $S_{Tot}$, $S_j^{P2P}$, and $S_{Tot}^{P2P}$ by sellers, initial prices are offered by the sellers and they are encrypted with the public key of sellers.

*(3) Price Feedback (Line 19):* Following the encryption of prices, the Buyers' Algorithm with Homomorphic Encryption (Algorithm 5) is

---

**Algorithm 5:** Buyers' Algorithm with HE

**Input** : $[\{\pi_1\}_{\mathcal{E}}, \{\pi_2\}_{\mathcal{E}}, \{\pi_3\}_{\mathcal{E}}, ...., \{\pi_{N_S}\}_{\mathcal{E}}]$
**Output**: $[A_{B_1}, ..., A_{B_{N_S}}], [\Omega_{B_1}, ..., \Omega_{B_{N_S}}], A_{Tot}, \Omega_{Tot}$

1 **for** Each $s_j$ **do**
2    **for** Each $b_i$ **do**
3       $\{-\pi_j\}_{\mathcal{E}} \leftarrow Smul(-1, \{\pi_j\}_{\mathcal{E}});$
4       $tmp1 \leftarrow Add(\lambda_i, \{-\pi_j\}_{\mathcal{E}});$
5       $\theta_{INV_i} \leftarrow \frac{1}{\theta_i};$
6       $\{X_{ji}\}_{\mathcal{E}} \leftarrow Smul(tmp1, \theta_{INV_i});$
7    **end**
8    $A_{B_j} \leftarrow 0;$
9    $\Omega_{B_j} \leftarrow [];$
10    **for** Each $b_i$ **do**
11       $\{\alpha_{ij}, \omega_{ij}\} \leftarrow Square(\{X_{ji}\}_{\mathcal{E}});$
12       $A_{B_j} \leftarrow A_{B_j} + Smul(0.5, Smul(\theta_i, \alpha_{ij}));$
13       $\Omega_{B_j} \leftarrow [\Omega_{B_j}||\{Smul(0.5, Smul(\theta_i, \omega_{ij}))\}];$
14    **end**
15 **end**
16 $A_{Tot} \leftarrow 0;$
17 $\Omega_{Tot} \leftarrow [];$
18 **for** Each $s_j$ **do**
19    $A_{Tot} \leftarrow A_{Tot} + A_{B_j};$
20    $\Omega_{Tot} \leftarrow [\Omega_{Tot}||\Omega_{B_j}];$
21 **end**

---

invoked, taking the input of encrypted prices, i.e. $[\{\pi_1\}_{\mathcal{E}}, ...., \{\pi_{N_S}\}_{\mathcal{E}}]$ (where the notation $\{\cdot\}_{\mathcal{E}}$ signifies that the enclosed data between curly braces is homomorphically encrypted). Within this algorithm, buyers compute their reaction functions over the encrypted data and subsequently return the outputs of their homomorphically encrypted calculations to the sellers.

It is important to note that when a special squaring operation is applied over homomorphically encrypted data, the outputs are represented as alphas and omegas, as described in Section 3.4.2. Therefore, the buyers transmit to the sellers the Alpha and Omega values representing $W_{B_j}$ for each $s_j$ and $W_{Tot}$ as output from Algorithm 5, i.e. $[A_{B_1}, ..., A_{B_{N_S}}], [\Omega_{B_1}, ..., \Omega_{B_{N_S}}], A_{Tot}, \Omega_{Tot}$.

*(4) Price Adjustment and Market Equilibrium (Lines 20–35):* Upon receiving the alpha and omega values ($[A_{B_1}, ..., A_{B_{N_S}}], [\Omega_{B_1}, ..., \Omega_{B_{N_S}}], A_{Tot}, \Omega_{Tot}$) from the Buyers' algorithm in line 19, the values of $W_{B_j}$ for each $s_j$ and $W_{Tot}$ are obtained by decrypting and processing alpha and omega values with specific operations between line 20 and line 29. The process begins with the direct decryption of the alpha value, setting it to the desired target value. Subsequently, each omega value is decrypted, squared, and cumulatively added to the target value. The decryption principles of encrypted data, which is involved in square operation, are explained in Section 3.4.2.

After obtaining $W_{B_j}$ for each $s_j$ and $W_{Tot}$ in plaintext format, the subsequent iteration's prices are computed between lines 30 and 34. This procedure mirrors the steps outlined between lines 19 and 23 in the sellers' algorithm (Algorithm 2), the details of which were previously explained in Section 4.1.1. Hence, to prevent redundancy, a detailed explanation of these steps is omitted here.

The aforementioned operations are reiterated in each iteration until the system attains market equilibrium, where the demand aligns with the supply — expressed as $|(D_j - S_j^{P2P})| < \epsilon$ for each $s_j$.

### 4.4. Buyers' algorithm with HE

The Buyers' Algorithm with HE (Algorithm 5) operates by processing encrypted prices and conducting all computations over homo-morphically encrypted data. It then produces the output in encrypted format.

This algorithm, a privacy-preserving adaptation of the Buyers' algorithm (Algorithm 3), ensures that all calculations are performed in a computationally efficient manner while protecting the privacy of the users. This is achieved through a computationally lightweight combination of primitive operations from the Paillier cryptosystem and the square operation detailed in Sections 3.4.1 and 3.4.2. By strategically avoiding the multiplication of two encrypted data sets, this approach circumvents the need for the more computationally demanding Fully Homomorphic Encryption.

The Buyers' algorithm includes the following tasks: (1) Calculating $X_{ji}$, representing the quantity of electricity buyer $b_i$ seeks to purchase from seller $s_j$. (2) Determining the Welfares for each seller $s_j$ and computing the Total Welfare.

Now, these tasks are explained in detail:

*(1) Calculating $X_{ji}$*: The $X_{ji}$ - amount of electricity that $b_i$ wishes to buy from $s_j$, is calculated in encrypted format ($\{X_{ji}\}_{\mathcal{E}}$) between line 3–6. In order to calculate $\{X_{ji}\}_{\mathcal{E}}$, first of all $\{-\pi_j\}_{\mathcal{E}}$ is calculated by multiplying $\{\pi_j\}_{\mathcal{E}}$ with "$-1$" by scalar multiplication in line 3. $\{-\pi_j\}_{\mathcal{E}}$ is added to $\lambda_i$ in line 4. The result of this operation is multiplied by the inverse of $\theta_i$ to obtain $\{X_{ji}\}_{\mathcal{E}}$ in line 5 and line 6. Please note that inverse operation in line 5 is performed individually on each $b_i$, just like the other operations conducted between lines 3 and 6. When executing these operations, the concluding values are generated in an encrypted format. This implies that the confidential user data, including $\lambda_i$ and $\theta_i$, remains private throughout the processing as the ultimate results of the equations are in the encrypted format before they are shared with other users.

*(2) Determining the Welfares*: the calculation of Eq. (10) is carried out for each seller in an encrypted format. Initially, the square of $\{X_{ji}\}_{\mathcal{E}}$ is computed in line 11, employing Algorithm 1 (Squaring). Then, square of $\{X_{ji}\}_{\mathcal{E}}$ is multiplied by $\theta_i$ and 1/2 respectively which are accumulated on $A_{B_j}$ and $\Omega_{B_j}$. Accumulation is done via addition on $A_{B_j}$ and via concatenation on $\Omega_{B_j}$. After this operation, Alpha and Omega values ($A_{B_j}, \Omega_{B_j}$) for each $s_j$ are accumulated on ($A_{Tot}, \Omega_{Tot}$) to calculate $W_{Tot}$ in encrypted format in lines 19 and 20. Finally, Buyers' algorithm returns back to Sellers' Algorithm (Algorithm 4) again with the Alpha and Omega values representing $W_{B_j}$ for each $s_j$ and $W_{Tot}$ as output, i.e. $[A_{B_1}, ..., A_{B_{N_S}}], [\Omega_{B_1}, ..., \Omega_{B_{N_S}}], A_{Tot}, \Omega_{Tot}$.

## 5. Evaluation

### 5.1. Privacy analysis

This section presents an analysis of the privacy guarantees provided by the encryption mechanisms employed in our system. We focus on proving the confidentiality of seller prices, demand data, and buyer profile variables.

#### 5.1.1. Seller price confidentiality

Seller prices undergo encryption using the seller's public key, denoted as $PK_S$, prior to transmission to buyers. Given that only sellers possess the corresponding private key of the sellers, represented as $SK_S$, buyers are consequently unable to discern the pricing details set by sellers.

Through the application of a chosen-plaintext attack (CPA) model, we demonstrate that seller prices remain confidential.

**Proof.** Let $PK_S$ be the public key used for encrypting seller prices, and $SK_S$ the corresponding private key. Assume an adversary $\mathcal{A}$ attempts to learn information about the homomorphically encrypted seller price $\text{Enc}_{pe}(PK_S, price)$.

(1) $\mathcal{A}$ chooses a price $price_1$ and obtains its encryption $\text{Enc}_{pe}(PK_S, price_1)$. (2) $\mathcal{A}$ is then given a challenge for the encryption

of either $price_1$ or another price $price_2$ chosen by a challenger. $\mathcal{A}$ must determine which price was encrypted.

A notable characteristic of the Paillier cryptosystem we employed is its provision for probabilistic encryption. Specifically, the system ensures that the encryption of identical plaintexts, when performed multiple times, yields distinct ciphertexts of a consistent length. This property significantly enhances the security of the encryption scheme by obfuscating the frequency of specific plaintext values within a corpus of encrypted data, thereby complicating attempts at cryptanalysis. The underlying mechanism for this feature is detailed in the foundational work by Paillier [65]. As a result, $\mathcal{A}$ cannot distinguish $Enc_{pe}(PK_S, price_1)$ from $Enc_{pe}(PK_S, price_2)$ with any significant advantage over random guessing. Hence, the confidentiality of seller prices is maintained.

### 5.1.2. Demand data confidentiality

The computation of the aggregate demand strategies by buyers for each seller is executed on data that has been encrypted, thus rendering it accessible exclusively through decryption with the private key, securely held by the sellers. This configuration guarantees that solely the sellers have the capability to access the consolidated information on final demand, thereby protecting this sensitive information from the buyers.

**Proof.** The preservation of confidentiality for the aggregated demand data is predicated on the exclusivity of decryption capabilities to the sellers.

Utilising an encryption protocol that is secure against chosen-plaintext attacks (CPA-secure), it is established that any adversary without access to the corresponding private key, represented as $SK_S$, is unable to decrypt or extract information from the encrypted demand strategy $Enc_{pe}(PK_S, demand)$. This framework effectively protects the confidentiality of demand-related data acquired through strategy function.

### 5.1.3. Buyer profile variables confidentiality

While buyer profile variables are encrypted with the seller's public key $(PK_S)$ for calculations on the buyer side, only the computed results are transmitted to the sellers. Consequently, sellers are unable to trace back individual buyer profile variables from the outcome of computations, ensuring the confidentiality of user-specific information.

**Proof.** Let us define a collection of buyer profile variables, denoted by $\{\lambda_1, \lambda_2, \ldots, \lambda_n\}$, which undergo encryption utilising the public key $PK_S$. The encrypted result $Enc_{pe}(PK_S, result)$ is computed based on these variables.

For the adversary $\mathcal{A}$, tracing back to different sets of buyer variables from aggregate result of operations is as hard as breaking the underlying encryption scheme's CPA security. Therefore, $\mathcal{A}$'s ability to infer any individual buyer profile variable from $Enc_{pe}(PK_S, result)$ is not possible, ensuring their confidentiality.

### 5.2. Existence of a nash equilibrium

In this subsection, we formally prove the existence of a Nash equilibrium for our model. In a Nash equilibrium, no user can improve its payoff by changing its strategies any more [9]. Thus, we check whether a Nash Equilibrium exists in the proposed work. The Nash equilibrium (NE) is a fundamental concept in game theory that defines a situation in which each player chooses the best strategy given the choices of the other players, and no individual can enhance their outcome unilaterally by changing their strategy [29,68].

If the following conditions are met in a game, then NE exists: (1) a finite number of players, (2) bounded, convex, and closed strategy sets, and (3) continuous and concave utility functions in the strategy space [69].

In our proposed model,

1. *player set is finite:* The number of participants in the player set is limited to sellers and buyers.

2. *strategy sets are bounded, convex and closed:* In our method, the game basically involves sellers and buyers, with sellers selecting their strategies based on the pricing they propose and buyers determining their strategies based on the amount of energy $X_{ji}$ they intend to buy from the sellers. The prices offered by sellers are subject to upper and lower bounds such that $\rho_{FiT} < \pi_j < \rho_{Sup}$, whereas the amount of energy $X_{ji}$ purchased by buyers is determined by a variety of parameters, including the proposed price $\pi_j$, $\lambda_i$, and $\theta_i$ which have upper and lower bounds or constant values implying that $X_{ji}$ has upper and lower limitations. As a result, the strategy sets formed by the proposed prices $\pi_j$ and purchased energy quantities $X_{ji}$ are non-empty, closed, bounded, and convex.

3. *utility functions are continuous and concave in the strategy space:* In our model, the buyers' goal is to maximise their utility function, $U_i$, by picking the optimum strategy, $X_{ji}$, depending on the sellers' proposed prices. The second derivative of the utility function $U_i$ with respect to the purchased energy quantity $X_{ji}$, is negative i.e,

$$\frac{\partial^2 U_i}{\partial x_{ji}^2} = -\theta_i < 0 \qquad (11)$$

so, $U_i$ is strictly concave in $X_{ji}$. Therefore, a Nash equilibrium exists in this game.

The analysis is valid for both homomorphically encrypted and non-encrypted formats because the same operations are carried out using the same data for both methods. Therefore, our approach is not dependent on the format used for data encryption with HE or non-encryption. This means that our findings can be applied to both scenarios.

### 5.3. Computational complexity

The Sellers' Algorithm is the main function of the Game Theoretical Energy Trade algorithm, which in turn invokes the Buyers' Method. The Sellers' Algorithm plays the position of the first mover in the Stackelberg Game, while the buyers are called the next movers. The time complexity of the method is determined by the Sellers' Algorithm's input parameters, which include the number of sellers and buyers [$N_S$, $N_B$]. The time required to perform the algorithm grows depending on the size of the input. Specifically, the dual nested loops starting from line 1 of Algorithm 3, line 20 of Algorithm 4, and line 1 (including the inner loops in lines 2 and 10) of Algorithm 5 contribute to a quadratic time complexity of $\mathcal{O}(n^2)$, with 'n' denoting the input size. Similarly, the loop and sum operation on line 5 of Algorithm 3 also results in a quadratic time complexity. The other parts of the algorithm have mostly linear or constant time complexity. Overall, the Game Theoretical Energy Trading method has a quadratic time complexity of $\mathcal{O}(n^2)$

### 5.4. Communication overhead

In this study, communication channels transmit exclusively homomorphically encrypted data, represented as $\{data\}_{\mathcal{E}}$. In every iteration of the Stackelberg Games, denoted by $N_{Iter}$, a total of $N_S \times N_B$ homomorphically encrypted price variables are transmitted from each seller to each buyer. Subsequently, $N_S \times N_B$ encrypted variables, representing the amount of energy each buyer intends to purchase from a seller, are transferred to the Aggregator.

The Aggregator processes the information by computing the welfare for each seller, $s_j$, denoted as $W_{B_j}$, as well as the total welfare $W_{Tot}$, all in a homomorphically encrypted format. The computation of $W_{B_j}$ for each seller $s_j$ produces one $A_{B_j}$ and $N_B$ concatenated encrypted values for $\Omega_{B_j}$. Thus, $1 + N_B$ encrypted values are output for each $W_{B_j}$, resulting in $N_S \times (1 + N_B)$ encrypted values for all sellers.

**Table 2**
Communication Overhead.

| Direction | Number of Bits |
|---|---|
| From each seller to each buyer | $N_{Iter} \times N_S \times N_B \times |\{data\}_{\mathcal{E}}|$ |
| From buyers to aggregator | $N_{Iter} \times N_S \times N_B \times |\{data\}_{\mathcal{E}}|$ |
| From aggregator to sellers | $N_{Iter} \times (2 \times N_S \times N_B + N_S + 1) \times |\{data\}_{\mathcal{E}}|$ |
| TOTAL | $N_{Iter} \times (4 \times N_S \times N_B + N_S + 1) \times |\{data\}_{\mathcal{E}}|$ |

The computation of $W_{Tot}$ in encrypted format produces one $A_{Tot}$ and $N_S \times N_B$ concatenated encrypted values for $\Omega_{B_{Tot}}$, so the aggregator outputs $1 + N_S \times N_B$ encrypted values for $W_{Tot}$. In total, the aggregator outputs $N_S \times (1 + N_B) + 1 + N_S \times N_B$ encrypted values, which simplifies to $2 \times N_S \times N_B + N_S + 1$.

Subsequently, these calculated values are sent back to the sellers in each iteration. In our implementation, we have set the length of homomorphically encrypted data to 4096 bits (i.e. $|\{data\}_{\mathcal{E}}| = 4096$ bits) The number of bits exchanged in each communication transfer is detailed in Table 2.

### 5.5. Experimental results

This section begins by detailing the data generation process for the experiments, followed by a comprehensive discussion of the simulation results.

#### 5.5.1. Data generation

The data generated for the experiments comprises household energy consumption and solar panel energy generation data. The code required for data generation is accessible on our GitHub repository.[1]

To generate household energy consumption data, we leverage electricity consumption data from 62 households in Germany, with each household representing a distinct socio-economic profile.[2] The energy consumption data, originally recorded in minutes, is aggregated and converted into hourly intervals.[3] Energy consumption data of 62 households are expanded to 300 households. To achieve this, the process involves randomly selecting one household out of a total of 62 households. For each selected household, data diversification is performed, and the results are subsequently stored as a newly generated household data. The diversification process entails the multiplication of each data point by a random factor uniformly distributed between 0.9 and 1.1. Through this process, the number of households expands with the addition of the newly generated ones, resulting in a total of 300 households. Within the dataset of generated household information, the initial 150 entries represent prosumer consumption data, while the subsequent 150 entries present consumer consumption data. The source code for household data generation and the generated energy consumption data is available in our repository.[4]

To generate solar panel (PV) electricity generation data, we employ the European Commission's photovoltaic geographical information system tool (PVGIS).[5] The specific parameters used for generating PV electricity generation data include the following: [*location*: Munich, Germany, *year*: 2016, *mounting type*: fixed, *Optimise slope and azimuth*: Yes *peak power*: 4KWp, *system loss*: 0]. We generated three distinct time-series data for solar panel electricity generation by combining Era 5 database with three different PV technologies (Crystalline silicon, CIS, and CdTe), all of which had their parameters specified as mentioned earlier. The generated time-series data can be found in our repository,[6] with hourly data available for each day of 2016.

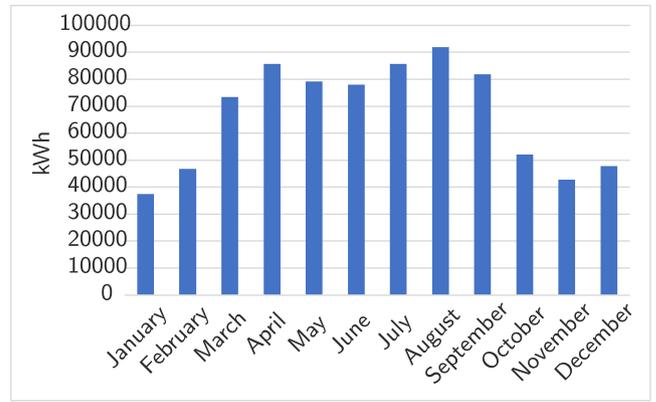

**Fig. 2.** Monthly total PV electricity generation in 2016.

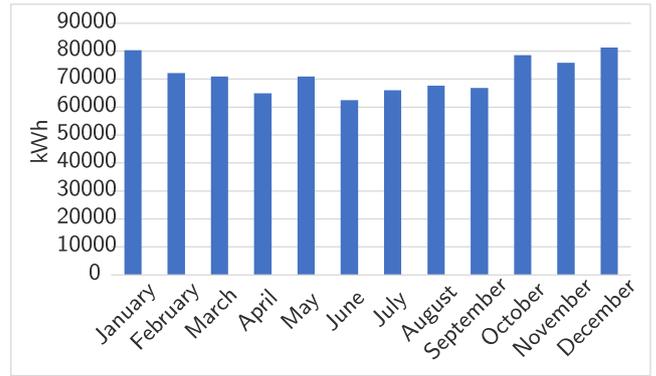

**Fig. 3.** Monthly total electricity consumption of households in 2016.

We utilise the three time-series, created using a PVGIS tool, to generate energy production data for 150 prosumers. This is achieved with an expansion process by randomly selecting a time series from the three generated ones and multiplying each data point in the randomly chosen time series by a random factor uniformly distributed between 0.9 and 1.1. This random process was done 150 times to cover all prosumer households. The codes for expansion and the generated data are available in our repository.[7]

The Fig. 2 illustrates the total monthly electricity generation from solar panels in the year 2016, as per the generated dataset. The graphical representation indicates that August witnessed the highest energy production, whereas January recorded the lowest among the depicted months.

The visual representation in Fig. 3 depicts the aggregate monthly electricity consumption of households in the year 2016, based on the generated dataset. The graph highlights higher energy consumption during the Autumn and Winter months, contrasting with lower consumption during the Spring and Summer months among the presented time periods.

Fig. 4 illustrates the monthly ratio between total photovoltaic (PV) electricity generation and the overall electricity consumption of households, utilising the generated dataset. The graphical representation reveals that during the months spanning March to September, the aggregate PV energy production surpasses the total energy consumption, while the opposite trend is observed for the remaining months. Notably, in 2016, August exhibits the highest ratio, while January records the lowest in terms of this comparative measure. Hence, our experimental focus is particularly directed towards the months of August and January, given their status as extreme cases in comparison to the other months.

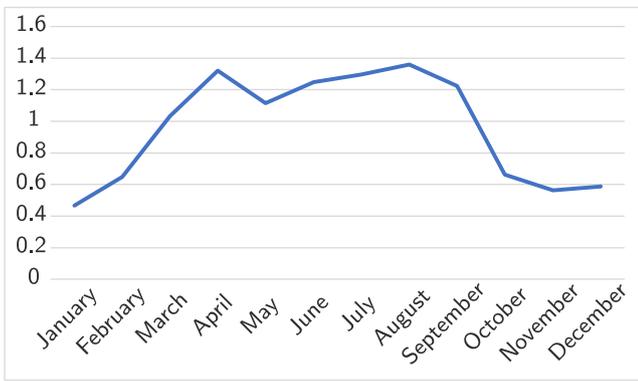

**Fig. 4.** The monthly ratio between total photovoltaic (PV) electricity generation and total electricity consumption of households in 2016.

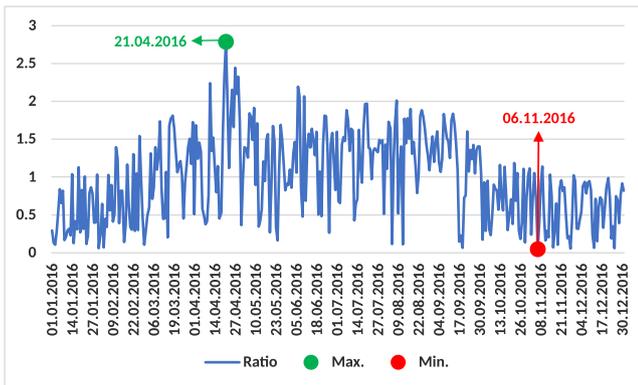

**Fig. 5.** The daily ratio between total photovoltaic (PV) electricity generation and total electricity consumption of households in 2016.

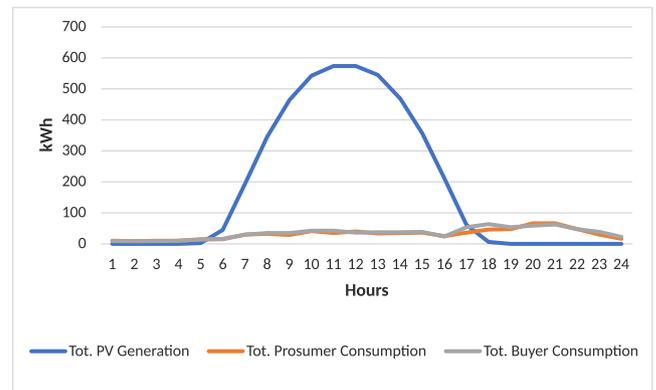

**Fig. 6.** Household electricity consumption and PV generation on April 21st, 2016.

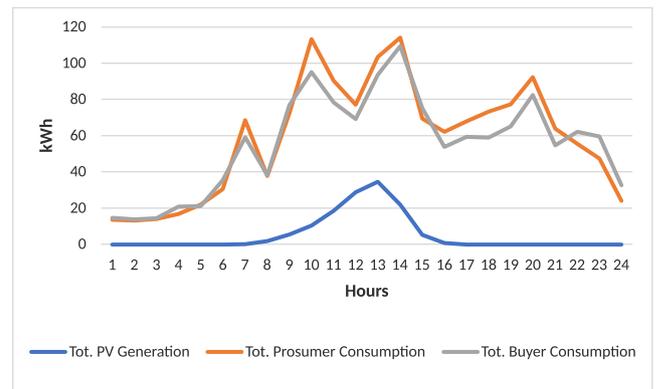

**Fig. 7.** Household electricity consumption and PV generation on November 6th, 2016.

Analysing the PV electricity generation dataset on a daily basis, as illustrated in Fig. 5 - depicting the ratio between total daily PV electricity generation and aggregate daily household electricity consumption in the year 2016 - we observe that the 21st of April holds the distinction of having the highest ratio of total PV electricity generation to total household electricity consumption. Similarly, the 6th of November also stands out with the lowest ratio in this context.

Given that the 21st of April and the 6th of November represent instances of pronounced extremities in energy generation and consumption, a deeper investigation into these specific dates is undertaken as follows.

The first analysis centres on the events of April 21st, a day marked by sunny weather favourable for effective electricity generation from PV panels. The relationship between high electricity generation and low energy consumption on this day is illustrated in Fig. 6.

Between 5 am and 6 pm, there is a notable surplus of energy generated, aligning with the peak sunlight hours that optimise PV panel performance. However, a significant observation emerges after 5 pm, as energy consumption noticeably increases. This surge in demand occurs just when solar energy production stops after 6 pm.

As a result, a critical issue arises — the excess energy generated before 6 pm becomes impractical for meeting the growing energy demand after 6 pm. Regrettably, this excess energy goes wasted, as it remains unused for the immediate energy consumption needs in the period following 6 pm. This inefficiency underscores a critical concern in the temporal alignment between energy production and consumption, thereby necessitating strategic interventions to mitigate such instances of energy wastage.

On November 6th, low energy generation from PV panels due to cloudy weather coincides with high energy consumption. This is driven by low outdoor temperatures prompting increased heating usage, compounded by indoor activities due to it being a Sunday with cold weather. Fig. 7 shows the disparity between energy generation and consumption, necessitating external energy supply.

Based on the aforementioned analyses, the experiments in the following subsection focus on August and January, with emphasis on April 21st and November 6th. These dates are pivotal for their extreme conditions. The experimentation covers the entire year 2016 for a comprehensive analysis.

### 5.5.2. Simulation results

In this part, the following simulation results are illustrated and explained.

*(1) Analysis of energy distributions:* The energy distribution within the energy market is depicted by examining various factors. This includes the consumption of energy by buyers from both P2P market and suppliers, the proportion of sellers among the prosumers, the quantity of excess energy injected into the P2P market by sellers, and the amount of excess energy consumed by sellers. Additionally, the energy consumption by prosumer consumers, sourced either through self-generation or from a supplier, is also highlighted. These simulations are performed for cases where users utilise a battery or not. In these simulations, we observe a notable increase in energy trading volume within the P2P market when utilising batteries. This serves as an effective solution to address the temporal misalignment between energy production and consumption, as depicted in Fig. 6.

**Table 3**
Distribution of energy within the market on April 21st, 2016.

| Ratio of Sellers | | | 25% | | | 50% | | | 75% | | |
|---|---|---|---|---|---|---|---|---|---|---|---|
| Number of Users | | | 40 | 120 | 200 | 40 | 120 | 200 | 40 | 120 | 200 |
| Without Battery | Buyer | Total from P2P | 70.01 | 216.60 | 378.77 | 58.52 | 175.38 | 256.07 | 34.51 | 78.25 | 143.10 |
| | | Total from supp. | 90.06 | 250.55 | 439.83 | 59.17 | 177.29 | 253.66 | 31.53 | 81.82 | 139.36 |
| | | Sellers % Ratio in Prosumers | 51.25 | 51.11 | 51.00 | 51.67 | 51.18 | 51.08 | 51.11 | 51.11 | 51.47 |
| | Sellers | Total to P2P | 70.01 | 216.60 | 378.77 | 58.51 | 175.38 | 256.07 | 34.51 | 78.25 | 143.10 |
| | | Total to grid. | 195.37 | 579.79 | 945.04 | 485.39 | 1417.00 | 2384.94 | 761.87 | 2299.84 | 3867.16 |
| | | Total self con. | 28.17 | 83.09 | 132.29 | 42.73 | 153.56 | 271.63 | 83.09 | 241.19 | 356.85 |
| | Cons. | Total from self gen. | 0.69 | 3.24 | 6.55 | 1.86 | 7.43 | 11.44 | 3.24 | 10.29 | 16.78 |
| | | Total from supp. | 23.07 | 70.01 | 132.15 | 38.18 | 152.12 | 263.29 | 70.01 | 231.09 | 380.48 |
| With Battery | Buyer | Total from P2P | 160.07 | 467.15 | 818.60 | 117.69 | 352.67 | 509.73 | 66.04 | 160.07 | 282.45 |
| | | Total from supp. | 0.00 | 0.00 | 0.00 | 0.00 | 0.00 | 0.00 | 0.00 | 0.00 | 0.00 |
| | | Sellers % Ratio in Prosumers | 95.42 | 97.22 | 97.75 | 100.00 | 100.00 | 100.00 | 100.00 | 100.00 | 100.00 |
| | Sellers | Total to P2P | 160.07 | 467.15 | 818.60 | 117.69 | 352.67 | 509.73 | 66.04 | 160.07 | 282.45 |
| | | Total to grid. | 24.69 | 112.50 | 152.58 | 347.30 | 990.16 | 1698.13 | 601.75 | 1885.60 | 3190.00 |
| | | Total self con. | 49.88 | 151.58 | 265.22 | 82.76 | 313.10 | 546.36 | 156.34 | 482.56 | 754.11 |
| | Cons. | Total from self gen. | 0.36 | 1.18 | 1.36 | 0.00 | 0.00 | 0.00 | 0.00 | 0.00 | 0.00 |
| | | Total from supp. | 1.69 | 3.58 | 4.41 | 0.00 | 0.00 | 0.00 | 0.00 | 0.00 | 0.00 |

**Table 4**
Distribution of energy within the market on November 6th, 2016.

| Ratio of Sellers | | | 25% | | | 50% | | | 75% | | |
|---|---|---|---|---|---|---|---|---|---|---|---|
| Number of Users | | | 40 | 120 | 200 | 40 | 120 | 200 | 40 | 120 | 200 |
| Without Battery | Buyer | Total from P2P | 1.75 | 6.40 | 12.38 | 3.57 | 14.18 | 22.33 | 6.40 | 20.28 | 30.45 |
| | | Total from supp. | 221.73 | 751.58 | 1333.72 | 161.07 | 505.46 | 824.82 | 87.52 | 203.20 | 416.90 |
| | | Sellers % Ratio in Prosumers | 10.00 | 10.28 | 11.25 | 9.58 | 10.76 | 10.13 | 10.28 | 10.28 | 9.47 |
| | Sellers | Total to P2P | 1.75 | 6.40 | 12.38 | 3.57 | 14.18 | 22.33 | 6.40 | 20.28 | 30.45 |
| | | Total to grid. | 0.00 | 0.00 | 0.00 | 0.00 | 0.00 | 0.00 | 0.00 | 0.00 | 0.00 |
| | | Total self con. | 2.21 | 5.50 | 8.14 | 4.42 | 10.04 | 16.11 | 5.50 | 15.16 | 24.22 |
| | Cons. | Total from self gen. | 4.11 | 12.78 | 22.36 | 9.03 | 27.23 | 46.95 | 12.78 | 41.97 | 73.43 |
| | | Total from supp. | 99.10 | 293.65 | 431.17 | 168.33 | 516.74 | 903.57 | 293.65 | 799.92 | 1324.30 |
| With Battery | Buyer | Total from P2P | 1.75 | 6.40 | 12.38 | 19.10 | 56.96 | 60.23 | 93.92 | 223.48 | 447.34 |
| | | Total from supp. | 221.73 | 751.58 | 1333.72 | 145.53 | 462.68 | 786.93 | 0.00 | 0.00 | 0.00 |
| | | Sellers % Ratio in Prosumers | 10.00 | 10.28 | 11.25 | 30.00 | 24.72 | 18.25 | 77.08 | 80.32 | 77.28 |
| | Sellers | Total to P2P | 1.75 | 6.40 | 12.38 | 19.10 | 56.96 | 60.23 | 93.92 | 223.48 | 447.34 |
| | | Total to grid. | 0.00 | 0.00 | 0.00 | 0.00 | 0.00 | 0.00 | 0.00 | 0.00 | 0.00 |
| | | Total self con. | 2.21 | 5.50 | 8.14 | 15.09 | 25.02 | 28.20 | 138.71 | 467.87 | 733.05 |
| | Cons. | Total from self gen. | 4.11 | 12.78 | 22.36 | 11.06 | 29.32 | 48.83 | 6.20 | 26.73 | 53.32 |
| | | Total from supp. | 99.10 | 293.65 | 431.17 | 155.64 | 499.67 | 889.60 | 167.02 | 362.45 | 635.58 |

*(2) Analysis of the costs, profits and balances:* The costs, profits and balances of buyers and prosumers within energy market is illustrated and analysed. Simulations are conducted for users with and without batteries. The scenarios cover the PP-LEM and PFET markets, along with the energy market excluding P2P interactions (No P2P). Each market scenario includes a results of buyer costs (from either the Supplier or P2P market), prosumer costs, prosumer profits (from either P2P or Feed-in Tariff), and the overall balance of all users. In these simulations, we conclude that PP-LEM and PFET markets emerge as a mechanism that enhances seller profits and lowers buyer costs, creating a compelling incentive structure and increasing the social welfare of the users. The choice between PFET and PPLEM P2P markets may yield varied profits or costs, yet user balances remain uncompromised.

*(3) Analysis of the dynamics of proposed Stackelberg Game:* Simulation results which illustrate the change of power prices and power demands over seller iterations are presented. In these simulations, we illustrate the pricing dynamics of sellers and demand dynamics of buyers, leveraging Stackelberg Game Theory principles.

*(4) Run-time performance analysis:* Run-time simulations showcasing the execution times of specific tasks for sellers and buyers in PP-LEM and PFET markets are presented. Our proposed solution, PP-LEM, showcases significantly enhanced computational efficiency in comparison to PFET [8].

The Simulation Parameters and Fixed prices used in our simulation are given as below:

**Simulation Parameters:** The following set of parameters is used in the simulations:

- $\eta_1 = 3$ which represents the step size employed to adjust the price of a seller in response to market demands for the subsequent iteration.
- $\eta_2 = 0.0001$ serves as the step size for updating states in the subsequent iteration. Specifically tailored for PFET, this small value is chosen to effectively scale high-order values to low-order states, ensuring a nuanced adjustment.
- $\lambda_n = 40.1$ represents the profile variable for $b_n$, characterising the behavioural aspects of prosumers in the system.
- $\theta = 25$ is a constant parameter, as specified in [66].

**Fixed Prices:** During the simulations, the following price values are used in alignment with the Germany market:

- **FiT:** is taken as 0.08 €/ kWh [70,71]
- **Supplier price:** is taken as 0.40 €/ kWh [72]

Now the results executed simulations are explained:

*(1) Analysis of energy distributions:* In Table 3, which depicts the energy market on April 21st, 2016, we observe dynamic energy interactions for scenarios where users either use a battery or do not. The upper half of the table, labelled "Without Battery", shows cases where users do not have batteries, while the lower half, labelled "With Battery", presents cases where users do have batteries. The ratio of prosumers in the table represents the proportion of prosumers among all users. The simulation results are run for the ratios of prosumers 25%, 50% and 75% and for the number of users 40, 120 and 200. From the upper half of the table, labelled "Without Battery", we can observe the amount of energy that buyers buy from the supplier is higher than the energy they buy from the P2P market. The sellers' ratio in prosumers shows that ratio of sellers in prosumers have covered all their self energy needs and they have excess electricity to sell. This part of the table, labelled "Without Battery" shows that only around 51% of the prosumers over the time is able to be a seller because they can only sell the electricity when the sun shines and PV panels produces electricity. As can be observed in Fig. 6, the prosumers have only surplus energy between 5 am and 6 pm. When the prosumers becomes sellers, the high amount of energy is injected into the grid. Lower amounts of energies are sold to P2P Market and lower amounts of energy are self consumed by the buyers. When the prosumers become consumers where they do not have excess electricity to sell, they buy most of their required energy as their produced energy is not sufficient to cover their needs. Although the prosumers produce huge amount of energy between 5 am and 6 pm, the surplus energy is not effectively utilised, because users mostly consume their energies after 5pm. There is a temporal misalignment between energy production and consumption. Due to this issue, buyers buy more energy from suppliers than from the P2P market, prosumer sellers inject most of their excess electricity into the grid rather than sell it to the P2P market, and prosumer consumers mostly cover their needed energies from the supplier. These observations indicate a higher dependence on suppliers compared to the P2P market.

The lower half of the Table 3, labelled "With Battery" which considers users equipped with batteries on April 21st, the P2P market undergoes significant changes. Thanks to the batteries, the high amount of surplus electricity that injected to the grid between 5 am and 6 pm, is stored in the batteries and the stored energy is used after 6pm, a time when energy is not produced but consumption is increased by the consumers. Thanks to the batteries, as illustrated in Table 3, buyers now exclusively fulfil all their energy requirements through P2P Markets, entirely bypassing reliance on suppliers. Sellers now exhibit a significantly higher proportion among prosumers, ranging from 95.42% to 100.00%. Consequently, prosumer sellers predominantly distribute their electricity through the P2P market rather than feeding it into the grid. This shift also indicates an enhanced self-sufficiency among prosumers in meeting their energy needs with their own generated electricity, thereby lowering reliance on suppliers to meet their demands.

In summary, the Tables 3 reveals the significant impact of user-equipped batteries on the P2P energy market. The integration of batteries significantly enhances the overall energy transactions in the P2P market. This shift enables buyers to fully leverage the cost-effectiveness of the P2P market, eliminating reliance on traditional suppliers entirely. Furthermore, the proportion of sellers among prosumers is poised to rise sharply, nearing 100%, allowing them to sell the majority of their surplus energy directly to the P2P market rather than feeding it back into the grid. Consequently, prosumers achieve greater self-sufficiency and better meet their energy needs independently.

The impact of battery technology on energy distribution within the market is also analysed for November 6, 2016, a date identified as having the lowest ratio of total energy production to total consumption throughout the year. The findings are presented in simulation results in Table 4. On this date, at a 25% Ratio of Prosumers (RoP), batteries were found to have minimal impact on enhancing the advantages for both buyers and prosumers due to a low number of sellers and insufficient energy production, preventing energy storage in batteries. However, at a 50% RoP, there was a slight improvement in the benefits for both buyers and prosumers, attributed to an increase in sellers. Most notably, at a 75% RoP, the benefits for buyers and prosumers saw significant improvements: buyers were able to purchase all their required energy from the P2P market, the proportion of sellers among prosumers rose to approximately 80%, prosumers increased their energy sales to the P2P market, relied more on self-generated energy, and, consequently, their dependence on traditional energy suppliers decreased.

The analysis of energy distribution across the market, taking into account users with and without batteries, is conducted through simulations for the entire months of August and January, as well as for the entire year of 2016. The results are presented in Table 8 for August, Table 9 for January, and Table 10 for the year 2016. Again, the data presented in these tables clearly demonstrates the benefits of utilising batteries for all user categories. With batteries, buyers can purchase more energy from the P2P market at lower prices compared to those offered by suppliers. Similarly, prosumer sellers can dispatch more energy to the P2P market, thereby achieving higher prices than those available through FiT. Furthermore, prosumers are capable of satisfying a greater portion of their energy needs through self-generation, reducing their dependence on external suppliers.

**(2) Analysis of the costs, profits and balances:** To enhance our understanding of the dynamics of the energy market, we examine the financial implications for users. Specifically, we compare their profits, costs, and balances across three distinct scenarios: a market without P2P trading (No P2P), a market utilising our previously proposed privacy-friendly energy trading platform (PFET) [8], and a market with our current proposal PP-LEM, where in PFET and PP-LEM markets, buyers can also purchase some part of their needed energies from P2P market, prosumers can sell some of their surplus energies to P2P market.

Table 5 presents an analysis of the financial outcomes for users on April 21st, 2016, across three market scenarios: No P2P, PFET, and PP-LEM, focusing specifically on situations where users do not utilise batteries or not. In the No P2P scenario, the exchange is straightforward: buyers obtain their required energy at the standard supplier rate, while prosumers sell their excess energy back to the grid at the Feed-in Tariff (FiT) rate. Should prosumers' self-generated energy be insufficient, they rely on purchasing additional energy from the supplier.

The introduction of PFET and PP-LEM markets brings notable financial benefits. In these scenarios, buyers are able to reduce their costs by purchasing energy through the P2P market at lower prices, and prosumers can increase their profits by selling excess energy at higher rates compared to the FiT price.

A closer examination of the PFET and PP-LEM markets reveals variations in costs and profits due to their distinct pricing policies. Nevertheless, it is important to note that the overall financial balances of users – calculated by subtracting total costs from total profits – remains consistent across both markets. This consistency is attributed to the P2P clearance mechanisms specific to each market, which only impact the direct costs and profits associated with P2P transactions. The mechanism ensures that the profits earned by prosumers are balanced by the costs incurred by buyers, neutralising any differences when calculating overall balances.

Comparing the financial balances in PFET and PP-LEM markets with the No P2P scenario highlights an improvement in user balances. This improvement is largely due to the P2P clearance mechanism, which proves advantageous for both prosumers and buyers by facilitating more efficient energy trades and enhancing financial returns for all parties involved.

Examining the data in Table 5, which compares user financial outcomes with and without the use of batteries, it is evident that incorporating batteries typically enhances user balances. When users achieve a 75% RoP, their balances see a significant boost, in some cases even doubling. At a 50% RoP, the impact varies by market scenario: in the No P2P scenario, there is a slight increase in balances, whereas in the PFET and PP-LEM scenarios, the increase is much more pronounced.

For a 25% RoP, both PFET and PP-LEM scenarios again exhibit a significant rise in user balances. Although it may initially appear that balances slightly decline in the No P2P scenario, this observation requires further clarification. The apparent decrease overlooks the fact that some energy remains stored in the batteries at the end of the day, intended for use the following day. This leftover energy is not included in the calculations for the current day's costs and profits, which might misleadingly suggest a decrease in balances. In reality,

**Table 5**
Profits and Costs within the energy market on April 21st, 2016.

| | | Ratio of Sellers | 25% | | | 50% | | | 75% | | |
|---|---|---|---|---|---|---|---|---|---|---|---|
| | | Number of Users | 40 | 120 | 200 | 40 | 120 | 200 | 40 | 120 | 200 |
| Without Battery | No P2P | Buyers Costs (Supp) | 64.03 | 186.86 | 327.44 | 47.08 | 141.07 | 203.89 | 26.42 | 64.03 | 112.98 |
| | | Prosumers Costs (Supp) | 9.23 | 28.00 | 52.86 | 15.27 | 60.85 | 105.32 | 28.00 | 92.44 | 152.19 |
| | | Prosumers Profits (FiT) | 21.23 | 63.71 | 105.90 | 43.51 | 127.39 | 211.28 | 63.71 | 190.25 | 320.82 |
| | | Prosumer Balance | 12.00 | 35.71 | 53.04 | 28.24 | 66.54 | 105.96 | 35.71 | 97.81 | 168.63 |
| | | All Users Balance | -52.03 | -151.15 | -274.40 | -18.84 | -74.52 | -97.93 | 9.29 | 33.78 | 55.65 |
| | PFET | Buyers Costs (P2P) | 24.04 | 74.53 | 128.73 | 19.31 | 57.86 | 86.43 | 10.63 | 25.64 | 46.09 |
| | | Buyers Costs (Supp) | 36.02 | 100.22 | 175.93 | 23.67 | 70.92 | 101.46 | 12.61 | 32.73 | 55.74 |
| | | Buyers Total Costs | 60.06 | 174.75 | 304.66 | 42.98 | 128.78 | 187.89 | 23.24 | 58.37 | 101.83 |
| | | Prosumers Costs | 9.23 | 28.00 | 52.86 | 15.27 | 60.85 | 105.32 | 28.00 | 92.44 | 152.19 |
| | | Prosumers Profits (P2P) | 24.04 | 74.53 | 128.73 | 19.31 | 57.86 | 86.43 | 10.63 | 25.64 | 46.09 |
| | | Prosumers Profits (FiT) | 15.63 | 46.38 | 75.60 | 38.83 | 113.36 | 190.80 | 60.95 | 183.99 | 309.37 |
| | | Prosumers Balance | 30.44 | 92.90 | 151.47 | 42.87 | 110.38 | 171.91 | 43.58 | 117.19 | 203.27 |
| | | All Users Balance | -29.62 | -81.84 | -153.19 | -0.11 | -18.40 | -15.98 | 20.33 | 58.82 | 101.44 |
| | PP-LEM | Buyers Costs (P2P) | 19.82 | 57.58 | 94.31 | 16.12 | 42.42 | 60.35 | 9.16 | 18.75 | 32.75 |
| | | Buyers Costs (Supp) | 36.02 | 100.22 | 175.93 | 23.67 | 70.92 | 101.46 | 12.61 | 32.73 | 55.74 |
| | | Buyers Total Costs | 55.84 | 157.80 | 270.24 | 39.79 | 113.34 | 161.81 | 21.77 | 51.47 | 88.49 |
| | | Prosumers Costs | 9.23 | 28.00 | 52.86 | 15.27 | 60.85 | 105.32 | 28.00 | 92.44 | 152.19 |
| | | Prosumers Profits (P2P) | 19.82 | 57.58 | 94.31 | 16.12 | 42.42 | 60.35 | 9.16 | 18.75 | 32.75 |
| | | Prosumers Profits (FiT) | 15.63 | 46.38 | 75.60 | 38.83 | 113.36 | 190.80 | 60.95 | 183.99 | 309.37 |
| | | Prosumers Balance | 26.22 | 75.96 | 117.05 | 39.68 | 94.93 | 145.83 | 42.11 | 110.30 | 189.93 |
| | | All Users Balance | -29.62 | -81.84 | -153.19 | -0.11 | -18.40 | -15.98 | 20.33 | 58.82 | 101.44 |
| With Battery | No P2P | Buyers Costs (Supp) | 64.03 | 186.86 | 327.44 | 47.08 | 141.07 | 203.89 | 26.42 | 64.03 | 112.98 |
| | | Prosumers Costs (Supp) | 0.67 | 1.41 | 1.75 | 0.00 | 0.00 | 0.00 | 0.00 | 0.00 | 0.00 |
| | | Prosumers Profits (FiT) | 8.07 | 27.87 | 45.33 | 33.00 | 94.91 | 158.75 | 51.17 | 157.77 | 267.83 |
| | | Prosumer Balance | 7.39 | 26.46 | 43.58 | 33.00 | 94.91 | 158.75 | 51.17 | 157.77 | 267.83 |
| | | All Users Balance | -56.64 | -160.40 | -283.86 | -14.07 | -46.15 | -45.14 | 24.75 | 93.75 | 154.85 |
| | PFET | Buyers Costs (P2P) | 52.25 | 154.30 | 267.58 | 37.78 | 112.12 | 166.69 | 19.60 | 50.59 | 88.03 |
| | | Buyers Costs (Supp) | 0.00 | 0.00 | 0.00 | 0.00 | 0.00 | 0.00 | 0.00 | 0.00 | 0.00 |
| | | Buyers Total Costs | 52.25 | 154.30 | 267.58 | 37.78 | 112.12 | 166.69 | 19.60 | 50.59 | 88.03 |
| | | Prosumers Costs | 0.67 | 1.43 | 1.77 | 0.00 | 0.00 | 0.00 | 0.00 | 0.00 | 0.00 |
| | | Prosumers Profits (P2P) | 52.25 | 154.30 | 267.58 | 37.78 | 112.12 | 166.69 | 19.60 | 50.59 | 88.03 |
| | | Prosumers Profits (FiT) | 1.98 | 9.00 | 12.21 | 27.78 | 79.21 | 135.85 | 48.14 | 150.85 | 255.20 |
| | | Prosumers Balance | 53.55 | 161.87 | 278.02 | 65.57 | 191.33 | 302.54 | 67.74 | 201.44 | 343.23 |
| | | All Users Balance | 1.30 | 7.57 | 10.44 | 27.78 | 79.21 | 135.85 | 48.14 | 150.85 | 255.20 |
| | PP-LEM | Buyers Costs (P2P) | 45.49 | 123.78 | 203.39 | 32.47 | 85.44 | 120.84 | 17.68 | 38.57 | 64.85 |
| | | Buyers Costs (Supp) | 0.00 | 0.00 | 0.00 | 0.00 | 0.00 | 0.00 | 0.00 | 0.00 | 0.00 |
| | | Buyers Total Costs | 45.49 | 123.78 | 203.39 | 32.47 | 85.44 | 120.84 | 17.68 | 38.57 | 64.85 |
| | | Prosumers Costs | 0.67 | 1.43 | 1.77 | 0.00 | 0.00 | 0.00 | 0.00 | 0.00 | 0.00 |
| | | Prosumers Profits (P2P) | 45.49 | 123.78 | 203.39 | 32.47 | 85.44 | 120.84 | 17.68 | 38.57 | 64.85 |
| | | Prosumers Profits (FiT) | 1.98 | 9.00 | 12.21 | 27.78 | 79.21 | 135.85 | 48.14 | 150.85 | 255.20 |
| | | Prosumers Balance | 46.79 | 131.35 | 213.83 | 60.25 | 164.65 | 256.69 | 65.82 | 189.42 | 320.05 |
| | | All Users Balance | 1.30 | 7.57 | 10.44 | 27.78 | 79.21 | 135.85 | 48.14 | 150.85 | 255.20 |

this stored energy represents a deferred asset that will contribute to the next days.

Our analysis extends to examining the costs and profits of users within energy markets on specific dates and periods, as detailed in Tables 11, 12, 13, and 14, covering November 6th, 2016, the entire months of August and January, and the full year of 2016, respectively. The insights derived from these tables mirror those observed in Table 5, highlighting consistent takeaways across different timescales. The key findings from our comprehensive cost and profit analysis include: (1) The use of batteries enhances user balances, demonstrating the financial benefits of energy storage. (2) Participation in P2P markets, such as PFET and PP-LEM, leads to increased user balances when compared to scenarios without P2P trading, underscoring the value of these market structures. (3) Balances of PFET and PP-LEM remain unchanged in different scenarios, indicating that the transition to PP-LEM from our earlier model, PFET, does not compromise the total balances and total social welfare of the users.

**(3) Analysis of the dynamics of proposed Stackelberg Game:** Now, we explore our proposed Stackelberg Game Theory-based clearance mechanism applied to PP-LEM, illustrating its dynamics and operational principles through simulation results illustrated in Figs. 8 and 9. The focus is on the interactions for a specific hour (12−13 pm) on April 21st, involving 40 users, of which 50% are prosumers. We examine the fluctuations in sellers' prices and changes in power demands directed at a particular seller across several game-theoretical iterations.

Figs. 8 and 9 showcase the variations in the prices offered by sellers and the power demands on one specific seller throughout the iterations for both PFET and PP-LEM simulations conducted on April 21st. These results highlight the negotiation process within a one-hour time slot, focusing on the initial ten sellers out of the total participants.

Reflecting on the system model outlined in Fig. 1, the process begins with sellers setting their initial prices. Based on these prices, demands for each seller are calculated. In response to the demand, sellers adjust their prices. This iterative process of price and demand adjustments continues until the supply and demand across the market align, demonstrating the dynamic interaction between buyers and sellers in the P2P energy trading mechanism. In this regard, PFET and PP-LEM operate under the same foundational principles, employing a same systematic approach. The key distinction between PFET and PP-LEM lies in the users' reaction functions — how users respond to their counterparts differs between these models. Specifically, consumers and sellers adjust their behaviour in unique ways: consumers calculate their demands based on the prices proposed by sellers differently, and sellers set their prices in response to the demands proposed by sellers in a distinct manner in each model. Therefore, due to the differences between the two models, both the final prices and the number of iterations required to reach market equilibrium vary between PFET and PP-LEM. It is worth noting that the working principles discussed for the time slot of 12−13 pm on April 21st, 2016, remain valid and consistent across other time slots and dates.

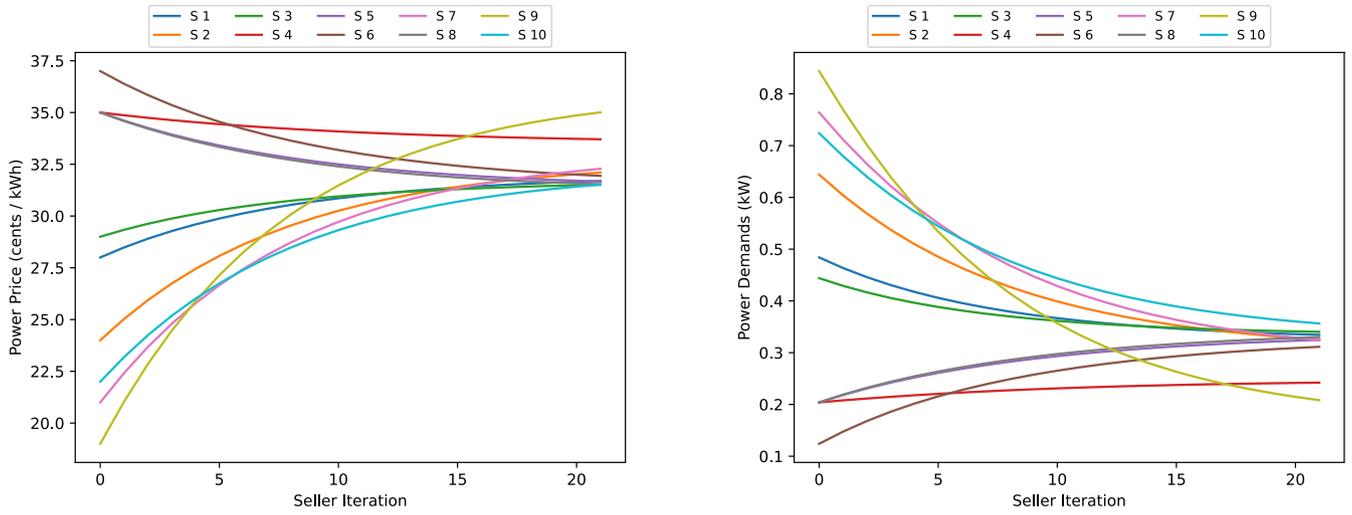

**Fig. 8.** Simulation results on 21st April for PFET.

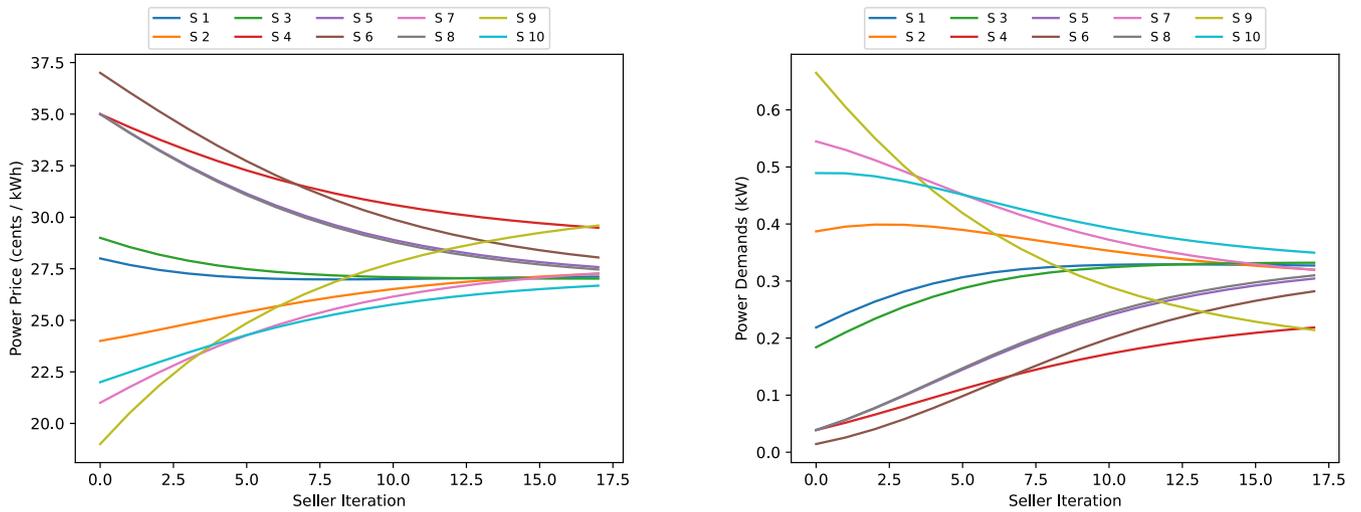

**Fig. 9.** Simulation results on 21st April for PP-LEM.

**Table 6**
Run-time Simulation Results.

| | | Ratio of prosumers | | | | | | | | | | | | | | |
| | | 25% | | | | | 50% | | | | | 75% | | | | |
| | Number of users | 40 | 80 | 120 | 160 | 200 | 40 | 80 | 120 | 160 | 200 | 40 | 80 | 120 | 160 | 200 |
|---|---|---|---|---|---|---|---|---|---|---|---|---|---|---|---|---|
| PFET [8] | Seller LC | 1.71 | 1.72 | 1.72 | 1.72 | 1.91 | 4.01 | 4.02 | 4.21 | 4.43 | 4.80 | 7.64 | 7.85 | 7.67 | 7.69 | 8.83 |
| | Buyer LC | 0.30 | 0.59 | 0.89 | 1.18 | 1.64 | 1.37 | 2.75 | 4.32 | 6.02 | 8.19 | 3.92 | 8.02 | 11.75 | 15.66 | 22.54 |
| | Buyer Agg. | 74.19 | 268.18 | 581.90 | 1014.12 | 1740.31 | 254.29 | 879.49 | 1963.30 | 3550.22 | 5959.85 | 458.04 | 1498.36 | 2964.60 | 5011.95 | 8783.70 |
| | Total | 76.20 | 270.49 | 584.51 | 1017.03 | 1743.87 | 259.67 | 886.26 | 1971.82 | 3560.67 | 5972.84 | 469.60 | 1514.23 | 2984.02 | 5035.30 | 8815.07 |
| | Num Iterations | 9 | 9 | 9 | 9 | 10 | 21 | 21 | 22 | 23 | 25 | 40 | 41 | 40 | 40 | 46 |
| | Average | **8.47** | **30.05** | **64.95** | **113.00** | **174.39** | **12.37** | **42.20** | **89.63** | **154.81** | **238.91** | **11.74** | **36.93** | **74.60** | **125.88** | **191.63** |
| PP-LEM | Seller LC | 8.82 | 17.73 | 26.38 | 35.64 | 47.33 | 10.46 | 26.78 | 42.96 | 58.41 | 77.63 | 11.13 | 24.96 | 35.62 | 49.26 | 65.80 |
| | Buyer LC | 0.67 | 1.39 | 2.00 | 2.66 | 3.62 | 1.90 | 5.31 | 9.02 | 12.51 | 16.92 | 5.07 | 13.61 | 21.07 | 29.43 | 40.95 |
| | Buyer Agg. | 14.89 | 65.74 | 155.41 | 276.90 | 469.00 | 34.36 | 191.22 | 486.37 | 898.11 | 1504.80 | 46.49 | 249.90 | 575.00 | 1069.15 | 1859.23 |
| | Total | 24.37 | 84.87 | 183.78 | 315.20 | 519.95 | 46.72 | 223.32 | 538.35 | 969.03 | 1599.35 | 62.69 | 288.48 | 631.69 | 1147.83 | 1965.98 |
| | Num Iterations | 9 | 10 | 12 | 12 | 13 | 18 | 25 | 28 | 29 | 31 | 32 | 43 | 44 | 46 | 51 |
| | Average | **2.71** | **8.49** | **15.32** | **26.27** | **40.00** | **2.60** | **8.93** | **19.23** | **33.41** | **51.59** | **1.96** | **6.71** | **14.36** | **24.95** | **38.55** |
| PFET/PP-LEM Cost | | **3.13** | **3.54** | **4.24** | **4.30** | **4.36** | **4.76** | **4.72** | **4.66** | **4.63** | **4.63** | **5.99** | **5.51** | **5.20** | **5.04** | **4.97** |

Let us further investigate Figs. 8 and 9. The initial energy supplies from these ten sellers to the peer-to-peer (P2P) market are recorded as follows: [0.323, 0.301, 0.332, 0.248, 0.336, 0.329, 0.292, 0.341, 0.165, and 0.328] kW. Concurrently, the initial prices they propose are [28, 24, 29, 35, 35, 37, 21, 35, 19, 22] euro cents, respectively. In the initial iteration, sellers set their prices, and based on these prices, users calculate the demands for each seller. For instance, at the start, the 9th seller offers a low price initially, while the 6th seller starts with a high price. At the initial iteration, as a reaction from consumers, demand is high for the 9th seller and low for the 6th seller, because there is an indirect correlation between prices offered and demand. In the next iterations individual sellers adjust their prices: lowering them if demand is low and increasing them if demand is high. The iteration, where sellers adjust prices and consumers adjust demands, concludes when the demand aligns with the supplies of the sellers. The termination condition for these iterations is that the gap between a seller's supply and demand should be no more than 0.05. As illustrated in Figs. 8 and 9, the PFET and PP-LEM result in distinct prices once the iteration concludes and the demands and supplies meet the stopping criteria. Additionally, the number of iterations required for both markets to reach this conclusion varies.

*(4) Run-time performance analysis:* Let us discuss the runtime simulations evaluating the performance of PFET and PP-LEM. The run-time results which are presented in Table 6 are conducted for a specific hour (12–13 pm) on April 21st, with a group of 40 users, half of whom are prosumers. In both PFET and PP-LEM, sellers perform their calculations independently and in parallel, while buyers also conduct some parts of their calculations independently and in parallel, but aggregation is not performed in parallel. In parallel processing, the slowest process is often called the "bottleneck," referring to a stage or component that limits overall performance because it takes longer to complete than other concurrent processes. In the parallel processing of price and demand calculations, the slowest component, known as the bottleneck, dictates the computation duration before transmitting values to subsequent stages. Table 6 presents Seller LC and Buyer LC, which sum up the longest computation durations for sellers and buyers in the parallel processing phase of each iteration. Buyer Agg. represents the total aggregation durations for buyers, while Total denotes the overall processing time. Average indicates the total run-time over number of iterations. PFET/PP-LEM Cost represents the ratio of average run-time costs between PFET and PP-LEM. As observed from Table 6, PP-LEM consistently exhibits shorter run-times compared to PFET. Examining PFET/PP-LEM Cost reveals that PP-LEM achieves significantly better computational efficiency, ranging from 3.13 to 5.99 times improvement. Clearly, PP-LEM demonstrates enhanced computational efficiency over PFET.

In the simulation section, we derive the following key insights:

1. Battery storage effectively addresses the temporal misalignment between energy generation and consumption, facilitating increased energy trading in P2P markets. Consequently, the integration of batteries enhances user balances.
2. Participation in Peer-to-Peer (P2P) markets, such as PFET and PP-LEM, results in increased user balances compared to scenarios without P2P trading. This underscores that these markets provide social welfare, particularly for users employing RES. Consequently, P2P trading incentivises the adoption of RES as a means to address the climate crisis.
3. Balances within PFET and PP-LEM remain stable across various scenarios, indicating that transitioning from our earlier model, PFET, to PP-LEM does not compromise the overall balances or social welfare of users.
4. PP-LEM exhibits markedly enhanced computational efficiency over PFET.

Based on these observations, we conclude that PP-LEM achieves exceptional computational enhancement while maintaining user balances and social welfare.

## 6. Comparison with existing privacy preserving clearance mechanisms

In this section, we compare our works (PP-LEM) with existing privacy-preserving clearance mechanisms, as outlined in Table 7. We compare PP-LEM with existing work based on privacy, decentralisation, data accuracy, scalability, and incentive competitive mechanisms for the game theoretical approach (ICMGT).

In terms of privacy, Zhang et al. [33], Li et al. [34], and Li et al. [35] have low levels of privacy protection due to reliance on anonymisation. Bevin et al. [39] has a medium level of privacy protection because the Market Operator has access to plaintext data of clusters, posing a risk of inferring household data. Xia et al. [49] also has a medium level of privacy protection due to reliance on anonymisation and desensitisation in inter-microgrid trading, as explained in Section 2.2 . The remaining works listed in Table 7 have high levels of privacy protection, thanks to the privacy-enhancing techniques they employ.

Regarding decentralisation, Zhang et al. [33], Li et al. [34], Li et al. [35], Sarenche et al. [36], Li et al. [37], Yang et al. [38], Li et al. [44], Hoseinpour et al. [47], and Li et al. [48] exhibit low decentralisation, as most calculations are performed by a single entity within the system. Bevin et al. [39] shows a medium level of decentralisation, with major double auction calculations performed by a market operator, although aggregation operations are performed by cluster coordinators. The remaining papers in Table 7 demonstrate high levels of decentralisation. Decentralisation is critical for security, as low decentralisation can lead to Single Points of Failure (SPF), reducing the system's security protection level, as detailed in [9].

In terms of data accuracy, Li et al. [44], Yu et al. [45], Hassan et al. [46], Hoseinpour et al. [47] and Li et al. [48] exhibit low levels of data accuracy due to their use of DP for protecting user privacy. The use of DP creates a trade-off between privacy and accuracy that can potentially lead to sub-optimal functionality.

Regarding scalability, Li et al. [37], Bevin et al. [39], Abidin et al. [40], Zobiri et al. [41], Zobiri et al. [42], Wang et al. [43], Hassan et al. [46], Hoseinpour et al. [47], Li et al. [48], Xia et al. [49], Erdayandi et al.[8] encounter notable challenges such that either they are not scalable in terms of computation or communication overhead, or the papers do not provide effective mechanisms to measure system performance, or do not report on it comprehensively.

Xie et al. [25] present a scalable system characterised by high levels of privacy, decentralisation, and data accuracy. However, a reported limitation of their work is the lack of incentive-competitive mechanisms within the game theoretical approach (ICMGT). In contrast, other game theoretical approaches, such as our previous work by Erdayandi et al. [8], address this issue by proposing an incentive-competitive mechanism. Despite this advancement, Erdayandi et al.'s system is not computationally efficient, which consequently leads to scalability issues.

To address these limitations, we propose PP-LEM, a competitive and scalable — computationally efficient mechanism. In Section 5.5.2, we conduct a comprehensive comparison of the computational efficiency of PP-LEM with that of Erdayandi et al. [8], demonstrating significant improvements in computational performance. PP-LEM not only enhances computational efficiency but also ensures privacy protection and preserves social welfare through an incentive clearance mechanism. Overall, our work introduces PP-LEM as a scalable — computationally efficient and competitive incentive mechanism that achieves high levels of privacy, decentralisation, and data accuracy.

## 7. Discussion

The main application of the proposed mechanism is to facilitate a clearance system for local energy markets. In this system, households with RES can sell their surplus electricity at prices higher than FiT, while households needing energy can purchase it at prices lower than those offered by suppliers. The system's main advantage is that it offers financial incentives to households using RES in a privacy-preserving and computationally efficient manner. It encourages the adoption of RES and supports broader efforts to combat climate change, fostering a sustainable and resilient energy future. The following subsections discuss potential practical cases of the mechanisms if the PP-LEM is generalised and describe the limitations of this work respectively.

### 7.1. Generalisation of PP-LEM and potential practical cases

The PP-LEM framework not only provides advantages for privacy-preserving clearance application for trading excess electricity from households, thereby encouraging the adoption of RES, but it can also be applied to various EV energy trading and demand flexibility scenarios.

**Table 7**
Comparison of This Work (PP-LEM) with Existing Privacy-Preserving Clearance Mechanisms.

| Paper | Year | Privacy | Decentralisation | Data Accuracy | Scalable | ICMGT |
|---|---|---|---|---|---|---|
| Zhang et al. [33] | 2017 | Low | Low | High | Yes | N/A |
| Li et al. [34] | 2018 | Low | Low | High | Yes | N/A |
| Li et al. [35] | 2018 | Low | Low | High | Yes | N/A |
| Sarenche et al. [36] | 2020 | High | Low | High | Yes | N/A |
| Li et al. [37] | 2017 | High | Low | High | NSLCE | N/A |
| Yang et al. [38] | 2020 | High | Low | High | Yes | N/A |
| Bevin et al. [39] | 2023 | Medium | Medium | High | NSLCE | N/A |
| Abidin et al. [40] | 2016 | High | High | High | NSLCE | N/A |
| Zobiri at al. [41] | 2022 | High | High | High | NSLCE | N/A |
| Zobiri at al. [42] | 2022 | High | High | High | NSLCE | N/A |
| Wang et al. [43] | 2023 | High | High | High | NSLCE | N/A |
| Li et al. [44] | 2019 | High | Low | Low | Yes | N/A |
| Yu et al. [45] | 2024 | High | High | Low | Yes | N/A |
| Hassan et al. [46] | 2019 | High | High | Low | NSLCE | N/A |
| Hoseinpour et al. [47] | 2023 | High | Low | Low | NSLCE | N/A |
| Li et al. [48] | 2023 | High | Low | Low | NSLCE | Yes |
| Xia et al. [49] | 2022 | Medium | High | High | NSLCE | Yes |
| Xie et al. [25] | 2020 | High | High | High | Yes | No |
| Erdayandi et al.[8] | 2022 | High | High | High | NSLCE | Yes |
| This Work (PP-LEM) | 2024 | High | High | High | Yes | Yes |

NSLCE: Not scalable or lacking comprehensive scalability evaluations, ICMGT: Incentive competitive mechanisms for the game theoretical approach, N/A: Not applicable.

**EV Energy Trading Application:** EV owners can store surplus energy for later use [73]. However, for those with solar panels, storage is constrained by the capacity of their EVs and household batteries [74], resulting in potential energy wastage once capacity is full [75]. EVs with surplus energy have the potential to participate in various energy trading activities [75]. They can sell their excess electricity to other EVs through vehicle-to-vehicle (V2V) energy trading [76], to the grid via vehicle-to-grid (V2G) mechanisms [77], or to residential homes through vehicle-to-house (V2H) trading [78]. The clearance mechanisms outlined in PP-LEM are also applicable for clearing electricity in V2V, V2G, and V2H scenarios. In addition, EV energy trading can expose sensitive data, such as locations, driving habits, and charging costs, and protecting this data is essential for privacy [79]. Privacy mechanisms in PP-LEM can be effectively utilised to protect user information in these applications.

**Demand Flexibility Application:** Within the framework of Demand Flexibility (DF) programmes, participants can adjust their electricity consumption patterns. Additionally, prosumer households with RES, such as solar panels, can generate surplus electricity and have the capability to sell this excess to other users. This strategy helps to flatten peak demands on suppliers by reacting to price signals or grid requirements [10]. The clearance methods outlined in PP-LEM can also be effectively utilised to manage surplus electricity in selling scenarios within DF programmes. Additionally, selling surplus energy involves sharing sensitive information about energy production and consumption, which necessitates robust data privacy protection measures [79]. Privacy mechanisms in PP-LEM can be effectively applied to clearance processes within DF programmes to protect user privacy.

### 7.2. Limitations of this work

In the PP-LEM, the costs associated with the distribution and transmission of electricity are disregarded. This simplification is practical in scenarios where the trading occurs within geographically compact communities. In such local contexts, the distances involved are minimal, and hence, the impact of distribution and transmission costs is negligible. However, when it comes to trading electricity between households that are located far apart, the situation changes significantly. In these cases, the network costs, which include both the distribution and transmission expenses, become a critical factor that must be taken into account. While Paudel et al. [80] propose a clearance mechanism that accounts for these network costs, it does not address privacy concerns. Conversely, the PP-LEM prioritises privacy but disregards distribution and transmission costs, highlighting a limitation. Thus, developing privacy-preserving clearance mechanisms that also incorporate network costs remains as an opportunity for future novel work.

## 8. Conclusion

The proposed PP-LEM application is a competitive Stackelberg Game clearance mechanism to create a privacy-preserving and computationally efficient energy trading market. This system, supported by homomorphic encryption, ensures the protection of sensitive information for both buyers and sellers. Our evaluation shows that the PP-LEM framework maintains user privacy more efficiently than current methods, achieving computational efficiency without compromising user welfare.

The main advantages of the proposed PP-LEM can be outlined as follows: (1) Financial incentives are provided to households using RES, increasing prosumer revenues and reducing consumer costs. (2) It promotes RES adoption, aiding in climate change efforts and supporting a sustainable energy future. (3) It performs clearance operations in a privacy-preserving and computationally efficient manner.

Key findings of this article are as follows: (1) Battery integration improves user balances. (2) Participation in P2P markets like PP-LEM increases user balances compared to non-P2P scenarios. (3) The clearance mechanism can be implemented in both a privacy-preserving and computationally efficient way. (4) PP-LEM offers better computational efficiency than state-of-the-art solutions like PFET without compromising social welfare.

The main disadvantage of this work is that PP-LEM does not account for distribution and transmission costs, limiting its applicability for communities with distant entities.

Future research directions include considering distribution and transmission costs in clearance mechanisms, exploring the balance between utility and privacy, with a focus on differential privacy techniques, and investigating cooperative techniques to enhance energy trading processes.

### CRediT authorship contribution statement

**Kamil Erdayandi:** Writing – review & editing, Writing – original draft, Visualization, Validation, Software, Resources, Methodology, Investigation, Formal analysis, Data curation, Conceptualization. **Mustafa A. Mustafa:** Writing – review & editing, Validation, Supervision, Project administration, Methodology, Investigation, Funding acquisition, Conceptualization.

**Table 8**
Distribution of energy within the market in August, 2016.

| | | Ratio of Sellers | 25% | | | 50% | | | 75% | | |
|---|---|---|---|---|---|---|---|---|---|---|---|
| | | Number of Users | 40 | 120 | 200 | 40 | 120 | 200 | 40 | 120 | 200 |
| Without Battery | Buyer | Total from P2P | 3004.42 | 8760.58 | 14966.80 | 2411.54 | 7086.03 | 11421.43 | 1425.82 | 3726.28 | 6354.00 |
| | | Total from supp. | 3962.07 | 11716.35 | 20326.18 | 2297.03 | 7156.57 | 11543.60 | 1215.57 | 3240.22 | 5716.29 |
| | | Sellers % Ratio in Prosumers | 45.40 | 44.61 | 45.53 | 45.80 | 45.70 | 45.64 | 44.61 | 45.76 | 45.74 |
| | Sellers | Total to P2P | 3004.42 | 8760.58 | 14966.80 | 2411.54 | 7086.03 | 11421.43 | 1425.82 | 3726.28 | 6354.00 |
| | | Total to grid | 1996.89 | 6057.69 | 10395.51 | 7893.83 | 23387.57 | 39006.14 | 13392.46 | 41847.16 | 69848.16 |
| | | Total self con. | 923.18 | 2870.22 | 4301.98 | 1632.86 | 5205.17 | 9027.87 | 2870.22 | 7947.63 | 13189.68 |
| | Cons. | Total from self gen. | 197.35 | 714.04 | 973.98 | 294.84 | 1088.35 | 1825.01 | 714.04 | 1609.66 | 2569.52 |
| | | Total from supp. | 1051.44 | 3580.22 | 5513.18 | 2031.63 | 6553.71 | 11215.26 | 3580.22 | 9945.94 | 16573.29 |
| With Battery | Buyer | Total from P2P | 4631.10 | 13529.72 | 23268.89 | 4708.56 | 14242.62 | 22965.03 | 2641.39 | 6966.49 | 12070.29 |
| | | Total from supp. | 2335.39 | 6947.22 | 12024.10 | 0.00 | 0.00 | 0.00 | 0.00 | 0.00 | 0.00 |
| | | Sellers % Ratio in Prosumers | 60.28 | 59.64 | 60.58 | 98.48 | 98.01 | 98.09 | 97.85 | 98.91 | 99.20 |
| | Sellers | Total to P2P | 4631.10 | 13529.72 | 23268.89 | 4708.56 | 14242.62 | 22965.03 | 2641.39 | 6966.49 | 12070.29 |
| | | Total to grid | 0.00 | 0.00 | 0.00 | 3598.55 | 9810.10 | 16337.53 | 8824.08 | 28925.58 | 47770.94 |
| | | Total self con. | 1380.27 | 4470.53 | 6798.11 | 3773.59 | 12212.98 | 21106.58 | 6785.66 | 18950.98 | 31716.03 |
| | Cons. | Total from self gen. | 110.47 | 402.29 | 571.28 | 55.04 | 165.84 | 259.53 | 94.46 | 134.98 | 144.38 |
| | | Total from supp. | 681.23 | 2291.73 | 3419.75 | 130.69 | 468.40 | 702.03 | 284.43 | 417.27 | 472.08 |

**Table 9**
Distribution of energy within the market in January, 2016.

| | | Ratio of Sellers | 25% | | | 50% | | | 75% | | |
|---|---|---|---|---|---|---|---|---|---|---|---|
| | | Number of Users | 40 | 120 | 200 | 40 | 120 | 200 | 40 | 120 | 200 |
| Without Battery | Buyer | Total from P2P | 1234.29 | 3761.05 | 6370.17 | 1312.32 | 3888.07 | 6322.45 | 947.44 | 2334.99 | 4135.31 |
| | | Total from supp. | 6331.74 | 20042.87 | 34377.58 | 3908.16 | 12500.02 | 20331.94 | 2135.13 | 5231.04 | 9693.76 |
| | | Sellers % Ratio in Prosumers | 24.02 | 23.85 | 24.40 | 25.19 | 24.37 | 24.05 | 23.85 | 24.16 | 24.22 |
| | Sellers | Total to P2P | 1234.29 | 3761.05 | 6370.17 | 1312.32 | 3888.07 | 6322.45 | 947.44 | 2334.99 | 4135.31 |
| | | Total to grid | 612.01 | 1679.96 | 2986.59 | 2584.82 | 7360.88 | 12171.71 | 4493.57 | 14392.87 | 23899.34 |
| | | Total self con. | 422.82 | 1275.14 | 1993.53 | 754.22 | 2395.31 | 4162.01 | 1275.14 | 3719.85 | 6131.66 |
| | Cons. | Total from self gen. | 238.70 | 767.91 | 1159.67 | 361.01 | 1358.48 | 2344.05 | 767.91 | 2078.83 | 3348.30 |
| | | Total from supp. | 2004.15 | 6377.96 | 10040.46 | 3572.33 | 11989.24 | 20478.62 | 6377.96 | 18230.89 | 30062.03 |
| With Battery | Buyer | Total from P2P | 1708.60 | 5064.30 | 8704.02 | 2821.26 | 8173.34 | 13193.90 | 2558.05 | 6573.62 | 11653.58 |
| | | Total from supp. | 5857.43 | 18739.62 | 32043.74 | 2399.23 | 8214.76 | 13460.49 | 524.53 | 992.41 | 2175.50 |
| | | Sellers % Ratio in Prosumers | 28.36 | 27.72 | 28.47 | 49.83 | 45.07 | 44.35 | 63.73 | 67.80 | 66.60 |
| | Sellers | Total to P2P | 1708.60 | 5064.30 | 8704.02 | 2821.26 | 8173.34 | 13193.90 | 2558.05 | 6573.62 | 11653.58 |
| | | Total to grid | 0.00 | 0.00 | 0.00 | 0.00 | 0.00 | 0.00 | 43.42 | 426.60 | 446.41 |
| | | Total self con. | 577.63 | 1702.03 | 2728.96 | 1857.05 | 5577.03 | 9635.66 | 4190.71 | 13626.85 | 22370.78 |
| | Cons. | Total from self gen. | 221.58 | 717.73 | 1076.98 | 334.06 | 1252.38 | 2170.65 | 599.11 | 1551.79 | 2578.09 |
| | | Total from supp. | 1866.46 | 6001.25 | 9387.72 | 2496.44 | 8913.62 | 15178.36 | 3631.17 | 8850.92 | 14593.13 |

**Table 10**
Distribution of energy within the market in 2016.

| | | Ratio of Sellers | 25% | | | 50% | | | 75% | | |
|---|---|---|---|---|---|---|---|---|---|---|---|
| | | Number of Users | 40 | 120 | 200 | 40 | 120 | 200 | 40 | 120 | 200 |
| Without Battery | Buyer | Total from P2P | 26123.98 | 79167.02 | 135348.02 | 23267.43 | 69687.58 | 113002.27 | 14555.47 | 36857.52 | 64892.76 |
| | | Total from supp. | 56296.47 | 176162.95 | 305454.15 | 33110.82 | 106312.84 | 173752.79 | 18220.24 | 45562.93 | 83217.55 |
| | | Sellers % Ratio in Prosumers | 37.49 | 37.30 | 37.91 | 38.59 | 37.99 | 37.77 | 37.30 | 37.84 | 37.94 |
| | Sellers | Total to P2P | 26123.98 | 79167.02 | 135348.02 | 23267.43 | 69687.58 | 113002.27 | 14555.47 | 36857.52 | 64892.76 |
| | | Total to grid | 16398.67 | 47429.38 | 80558.01 | 65151.16 | 189702.81 | 315182.01 | 112040.94 | 349755.28 | 582188.42 |
| | | Total self con. | 8534.08 | 25498.08 | 39645.64 | 15023.05 | 47868.67 | 83264.59 | 25498.08 | 74049.79 | 122088.45 |
| | Cons. | Total from self gen. | 2839.31 | 9317.90 | 13422.61 | 4209.18 | 15422.88 | 26410.89 | 9317.90 | 23458.59 | 37533.60 |
| | | Total from supp. | 16792.98 | 54479.11 | 85985.15 | 30673.14 | 102431.11 | 174969.48 | 54479.11 | 156284.82 | 257735.00 |
| With Battery | Buyer | Total from P2P | 39089.05 | 116218.58 | 199051.55 | 47334.00 | 143445.82 | 232425.51 | 31739.10 | 80685.71 | 143837.33 |
| | | Total from supp. | 43331.40 | 139111.39 | 241750.62 | 9044.25 | 32554.60 | 54329.55 | 1036.60 | 1734.74 | 4272.98 |
| | | Sellers % Ratio in Prosumers | 48.23 | 47.18 | 47.69 | 80.86 | 77.59 | 77.21 | 88.37 | 90.73 | 90.57 |
| | Sellers | Total to P2P | 39089.05 | 116218.58 | 199051.55 | 47334.00 | 143445.82 | 232425.51 | 31739.10 | 80685.71 | 143837.33 |
| | | Total to grid | 62.63 | 253.86 | 353.03 | 21484.02 | 55839.19 | 91655.99 | 56051.19 | 183089.58 | 297307.87 |
| | | Total self con. | 12502.99 | 37600.38 | 58866.59 | 36744.55 | 115578.02 | 200654.75 | 69754.25 | 211967.49 | 352606.74 |
| | Cons. | Total from self gen. | 2241.38 | 7339.55 | 10703.11 | 2013.81 | 7772.02 | 13047.24 | 3520.09 | 7318.75 | 11229.15 |
| | | Total from supp. | 13422.01 | 44355.15 | 69483.70 | 11147.01 | 42372.62 | 70942.96 | 16020.74 | 34506.97 | 53521.16 |

## Declaration of competing interest

The authors declare that they have no known competing financial interests or personal relationships that could have appeared to influence the work reported in this paper.

## Data availability

Data will be made available on request.

## Acknowledgement

We express our gratitude to Yi Dong for providing recommendations on solar panel electricity generation data and for providing valuable feedback.

## Appendix. Tables of experimental results

See Tables 8–14.

**Table 11**
Profits and Costs within energy market on November 6th, 2016.

| | | Ratio of Sellers | 25% | | | 50% | | | 75% | | |
|---|---|---|---|---|---|---|---|---|---|---|---|
| | | Number of Users | 40 | 120 | 200 | 40 | 120 | 200 | 40 | 120 | 200 |
| Without Battery | No P2P | Buyers Costs (Supp) | 89.39 | 303.19 | 538.44 | 65.85 | 207.86 | 338.86 | 37.57 | 89.39 | 178.94 |
| | | Prosumers Costs (Supp) | 39.64 | 117.46 | 172.47 | 67.33 | 206.70 | 361.43 | 117.46 | 319.97 | 529.72 |
| | | Prosumers Profits (FiT) | 0.14 | 0.51 | 0.99 | 0.29 | 1.13 | 1.79 | 0.51 | 1.62 | 2.44 |
| | | Prosumer Balance | -39.50 | -116.95 | -171.48 | -67.05 | -205.56 | -359.64 | -116.95 | -318.35 | -527.29 |
| | | All Users Balance | -128.89 | -420.14 | -709.92 | -132.90 | -413.42 | -698.50 | -154.51 | -407.74 | -706.22 |
| | PFET | Buyers Costs (P2P) | 0.69 | 2.52 | 4.88 | 1.35 | 5.37 | 8.48 | 2.15 | 6.88 | 10.43 |
| | | Buyers Costs (Supp) | 88.69 | 300.63 | 533.49 | 64.43 | 202.18 | 329.93 | 35.01 | 81.28 | 166.76 |
| | | Buyers Total Costs | 89.38 | 303.16 | 538.37 | 65.78 | 207.56 | 338.41 | 37.16 | 88.15 | 177.19 |
| | | Prosumers Costs | 39.64 | 117.46 | 172.47 | 67.33 | 206.70 | 361.43 | 117.46 | 319.97 | 529.72 |
| | | Prosumers Profits (P2P) | 0.69 | 2.52 | 4.88 | 1.35 | 5.37 | 8.48 | 2.15 | 6.88 | 10.43 |
| | | Prosumers Profits (FiT) | 0.00 | 0.00 | 0.00 | 0.00 | 0.00 | 0.00 | 0.00 | 0.00 | 0.00 |
| | | Prosumers Balance | -38.95 | -114.94 | -167.59 | -65.98 | -201.32 | -352.95 | -115.30 | -313.09 | -519.29 |
| | | All Users Balance | -128.34 | -418.09 | -705.96 | -131.76 | -408.88 | -691.36 | -152.47 | -401.25 | -696.48 |
| | PP-LEM | Buyers Costs (P2P) | 0.51 | 1.65 | 3.08 | 0.94 | 3.44 | 5.19 | 1.65 | 4.65 | 6.82 |
| | | Buyers Costs (Supp) | 88.69 | 300.63 | 533.49 | 64.43 | 202.18 | 329.93 | 35.01 | 81.28 | 166.76 |
| | | Buyers Total Costs | 89.20 | 302.28 | 536.57 | 65.36 | 205.62 | 335.12 | 36.66 | 85.93 | 173.58 |
| | | Prosumers Costs | 39.64 | 117.46 | 172.47 | 67.33 | 206.70 | 361.43 | 117.46 | 319.97 | 529.72 |
| | | Prosumers Profits (P2P) | 0.51 | 1.65 | 3.08 | 0.94 | 3.44 | 5.19 | 1.65 | 4.65 | 6.82 |
| | | Prosumers Profits (FiT) | 0.00 | 0.00 | 0.00 | 0.00 | 0.00 | 0.00 | 0.00 | 0.00 | 0.00 |
| | | Prosumers Balance | -39.13 | -115.81 | -169.39 | -66.40 | -203.26 | -356.24 | -115.81 | -315.32 | -522.90 |
| | | All Users Balance | -128.34 | -418.09 | -705.96 | -131.76 | -408.88 | -691.36 | -152.47 | -401.25 | -696.48 |
| With Battery | No P2P | Buyers Costs (Supp) | 89.39 | 303.19 | 538.44 | 65.85 | 207.86 | 338.86 | 37.57 | 89.39 | 178.94 |
| | | Prosumers Costs (Supp) | 38.94 | 115.53 | 169.09 | 55.50 | 185.92 | 341.63 | 64.84 | 140.26 | 244.21 |
| | | Prosumers Profits (FiT) | 0.00 | 0.00 | 0.00 | 0.00 | 0.00 | 0.00 | 0.00 | 0.00 | 0.00 |
| | | Prosumer Balance | -38.94 | -115.53 | -169.09 | -55.50 | -185.92 | -341.63 | -64.84 | -140.26 | -244.21 |
| | | All Users Balance | -128.34 | -418.73 | -707.54 | -121.35 | -393.78 | -680.49 | -102.41 | -229.65 | -423.15 |
| | PFET | Buyers Costs (P2P) | 0.69 | 2.50 | 4.87 | 6.88 | 20.52 | 22.25 | 22.96 | 63.58 | 122.54 |
| | | Buyers Costs (Supp) | 88.69 | 300.63 | 533.49 | 58.21 | 185.07 | 314.77 | 0.00 | 0.00 | 0.00 |
| | | Buyers Total Costs | 89.38 | 303.14 | 538.36 | 65.10 | 205.59 | 337.02 | 22.96 | 63.58 | 122.54 |
| | | Prosumers Costs | 39.64 | 117.46 | 172.47 | 62.25 | 199.87 | 355.84 | 66.81 | 144.98 | 254.23 |
| | | Prosumers Profits (P2P) | 0.69 | 2.50 | 4.87 | 6.88 | 20.52 | 22.25 | 22.96 | 63.58 | 122.54 |
| | | Prosumers Profits (FiT) | 0.00 | 0.00 | 0.00 | 0.00 | 0.00 | 0.00 | 0.00 | 0.00 | 0.00 |
| | | Prosumers Balance | -38.96 | -114.95 | -167.59 | -55.37 | -179.35 | -333.59 | -43.85 | -81.41 | -131.70 |
| | | All Users Balance | -128.34 | -418.09 | -705.96 | -120.47 | -384.94 | -670.61 | -66.81 | -144.98 | -254.23 |
| | PP-LEM | Buyers Costs (P2P) | 0.51 | 1.65 | 3.08 | 5.06 | 13.47 | 13.20 | 24.74 | 52.06 | 98.61 |
| | | Buyers Costs (Supp) | 88.69 | 300.63 | 533.49 | 58.21 | 185.07 | 314.77 | 0.00 | 0.00 | 0.00 |
| | | Buyers Total Costs | 89.20 | 302.28 | 536.57 | 63.27 | 198.54 | 327.97 | 24.74 | 52.06 | 98.61 |
| | | Prosumers Costs | 39.64 | 117.46 | 172.47 | 62.25 | 199.87 | 355.84 | 66.81 | 144.98 | 254.23 |
| | | Prosumers Profits (P2P) | 0.51 | 1.65 | 3.08 | 5.06 | 13.47 | 13.20 | 24.74 | 52.06 | 98.61 |
| | | Prosumers Profits (FiT) | 0.00 | 0.00 | 0.00 | 0.00 | 0.00 | 0.00 | 0.00 | 0.00 | 0.00 |
| | | Prosumers Balance | -39.13 | -115.81 | -169.39 | -57.20 | -186.40 | -342.64 | -42.07 | -92.93 | -155.62 |
| | | All Users Balance | -128.34 | -418.09 | -705.96 | -120.47 | -384.94 | -670.61 | -66.81 | -144.98 | -254.23 |

**Table 12**
Profits and Costs within energy market in August, 2016.

| | | Ratio of Sellers | 25% | | | 50% | | | 75% | | |
|---|---|---|---|---|---|---|---|---|---|---|---|
| | | Number of Users | 40 | 120 | 200 | 40 | 120 | 200 | 40 | 120 | 200 |
| Without Battery | No P2P | Buyers Costs (Supp) | 2786.60 | 8190.77 | 14117.19 | 1883.42 | 5697.04 | 9186.01 | 1056.56 | 2786.60 | 4828.12 |
| | | Prosumers Costs (Supp) | 420.57 | 1432.12 | 2205.27 | 812.65 | 2621.48 | 4486.10 | 1432.12 | 3978.37 | 6629.32 |
| | | Prosumers Profits (FiT) | 400.11 | 1185.46 | 2028.99 | 824.43 | 2437.89 | 4034.21 | 1185.46 | 3645.87 | 6096.17 |
| | | Prosumer Balance | -20.47 | -246.65 | -176.29 | 11.78 | -183.59 | -451.90 | -246.65 | -332.50 | -533.14 |
| | | All Users Balance | -2807.07 | -8437.43 | -14293.48 | -1871.65 | -5880.64 | -9637.91 | -1303.21 | -3119.10 | -5361.26 |
| | PFET | Buyers Costs (P2P) | 898.10 | 2662.49 | 4539.46 | 694.66 | 2075.08 | 3394.59 | 378.45 | 1073.18 | 1818.01 |
| | | Buyers Costs (Supp) | 1584.83 | 4686.54 | 8130.47 | 918.81 | 2862.63 | 4617.44 | 486.23 | 1296.09 | 2286.52 |
| | | Buyers Total Costs | 2482.93 | 7349.03 | 12669.93 | 1613.47 | 4937.71 | 8012.03 | 864.68 | 2369.27 | 4104.53 |
| | | Prosumers Costs | 420.57 | 1432.12 | 2205.27 | 812.65 | 2621.48 | 4486.10 | 1432.12 | 3978.37 | 6629.32 |
| | | Prosumers Profits (P2P) | 898.10 | 2662.49 | 4539.46 | 694.66 | 2075.08 | 3394.59 | 378.45 | 1073.18 | 1818.01 |
| | | Prosumers Profits (FiT) | 159.75 | 484.62 | 831.64 | 631.51 | 1871.01 | 3120.49 | 1071.40 | 3347.77 | 5587.85 |
| | | Prosumers Balance | 637.28 | 1714.98 | 3165.83 | 513.51 | 1324.60 | 2028.98 | 17.74 | 442.58 | 776.55 |
| | | All Users Balance | -1845.65 | -5634.04 | -9504.10 | -1099.95 | -3613.11 | -5983.05 | -846.95 | -1926.69 | -3327.98 |
| | PP-LEM | Buyers Costs (P2P) | 852.72 | 2315.40 | 3682.07 | 659.31 | 1703.80 | 2663.82 | 377.94 | 887.02 | 1444.69 |
| | | Buyers Costs (Supp) | 1584.83 | 4686.54 | 8130.47 | 918.81 | 2862.63 | 4617.44 | 486.23 | 1296.09 | 2286.52 |
| | | Buyers Total Costs | 2437.55 | 7001.94 | 11812.54 | 1578.12 | 4566.43 | 7281.26 | 864.16 | 2183.11 | 3731.21 |
| | | Prosumers Costs | 420.57 | 1432.12 | 2205.27 | 812.65 | 2621.48 | 4486.10 | 1432.12 | 3978.37 | 6629.32 |
| | | Prosumers Profits (P2P) | 852.72 | 2315.40 | 3682.07 | 659.31 | 1703.80 | 2663.82 | 377.94 | 887.02 | 1444.69 |
| | | Prosumers Profits (FiT) | 159.75 | 484.62 | 831.64 | 631.51 | 1871.01 | 3120.49 | 1071.40 | 3347.77 | 5587.85 |
| | | Prosumers Balance | 591.90 | 1367.90 | 2308.44 | 478.16 | 953.32 | 1298.21 | 17.22 | 256.42 | 403.23 |
| | | All Users Balance | -1845.65 | -5634.04 | -9504.10 | -1099.95 | -3613.11 | -5983.05 | -846.95 | -1926.69 | -3327.98 |
| With Battery | No P2P | Buyers Costs (Supp) | 2786.60 | 8190.77 | 14117.19 | 1883.42 | 5697.04 | 9186.01 | 1056.56 | 2786.60 | 4828.12 |
| | | Prosumers Costs (Supp) | 33.89 | 97.15 | 106.41 | 32.92 | 99.61 | 139.18 | 91.02 | 133.66 | 143.17 |
| | | Prosumers Profits (FiT) | 309.32 | 879.21 | 1542.58 | 654.33 | 1888.39 | 3087.60 | 908.28 | 2854.99 | 4762.09 |
| | | Prosumer Balance | 275.43 | 782.06 | 1436.17 | 621.41 | 1788.78 | 2948.42 | 817.26 | 2721.32 | 4618.92 |
| | | All Users Balance | -2511.17 | -7408.71 | -12681.02 | -1262.02 | -3908.26 | -6237.59 | -239.29 | -65.27 | -209.20 |
| | PFET | Buyers Costs (P2P) | 1330.24 | 3986.46 | 6818.09 | 1364.98 | 4144.35 | 6777.26 | 697.59 | 1989.01 | 3410.42 |
| | | Buyers Costs (Supp) | 934.16 | 2778.89 | 4809.64 | 0.00 | 0.00 | 0.00 | 0.00 | 0.00 | 0.00 |
| | | Buyers Total Costs | 2264.39 | 6765.35 | 11627.73 | 1364.98 | 4144.35 | 6777.26 | 697.59 | 1989.01 | 3410.42 |
| | | Prosumers Costs | 272.49 | 916.69 | 1367.90 | 52.28 | 187.36 | 280.81 | 113.77 | 166.91 | 188.83 |
| | | Prosumers Profits (P2P) | 1330.24 | 3986.46 | 6818.09 | 1364.98 | 4144.35 | 6777.26 | 697.59 | 1989.01 | 3410.42 |
| | | Prosumers Profits (FiT) | 0.00 | 0.00 | 0.00 | 287.88 | 784.81 | 1307.00 | 705.93 | 2314.05 | 3821.67 |
| | | Prosumers Balance | 1057.75 | 3069.77 | 5450.19 | 1600.59 | 4741.80 | 7803.45 | 1289.74 | 4136.15 | 7043.27 |
| | | All Users Balance | -1206.65 | -3695.58 | -6177.54 | 235.61 | 597.45 | 1026.19 | 592.16 | 2147.14 | 3632.84 |
| | PP-LEM | Buyers Costs (P2P) | 1310.04 | 3559.48 | 5713.90 | 1289.95 | 3407.18 | 5357.14 | 700.99 | 1666.76 | 2753.69 |
| | | Buyers Costs (Supp) | 934.16 | 2778.89 | 4809.64 | 0.00 | 0.00 | 0.00 | 0.00 | 0.00 | 0.00 |
| | | Buyers Total Costs | 2244.19 | 6338.36 | 10523.54 | 1289.95 | 3407.18 | 5357.14 | 700.99 | 1666.76 | 2753.69 |
| | | Prosumers Costs | 272.49 | 916.69 | 1367.90 | 52.28 | 187.36 | 280.81 | 113.77 | 166.91 | 188.83 |
| | | Prosumers Profits (P2P) | 1310.04 | 3559.48 | 5713.90 | 1289.95 | 3407.18 | 5357.14 | 700.99 | 1666.76 | 2753.69 |
| | | Prosumers Profits (FiT) | 0.00 | 0.00 | 0.00 | 287.88 | 784.81 | 1307.00 | 705.93 | 2314.05 | 3821.67 |
| | | Prosumers Balance | 1037.55 | 2642.78 | 4346.00 | 1525.56 | 4004.63 | 6383.33 | 1293.14 | 3813.90 | 6386.53 |
| | | All Users Balance | -1206.65 | -3695.58 | -6177.54 | 235.61 | 597.45 | 1026.19 | 592.16 | 2147.14 | 3632.84 |

**Table 13**
Profits and Costs within energy market in January, 2016.

| | | Ratio of Sellers | 25% | | | 50% | | | 75% | | |
|---|---|---|---|---|---|---|---|---|---|---|---|
| | | Number of Users | 40 | 120 | 200 | 40 | 120 | 200 | 40 | 120 | 200 |
| Without Battery | No P2P | Buyers Costs (Supp) | 3026.41 | 9521.57 | 16299.10 | 2088.19 | 6555.24 | 10661.75 | 1233.03 | 3026.41 | 5531.63 |
| | | Prosumers Costs (Supp) | 801.66 | 2551.18 | 4016.18 | 1428.93 | 4795.70 | 8191.45 | 2551.18 | 7292.35 | 12024.81 |
| | | Prosumers Profits (FiT) | 147.70 | 435.28 | 748.54 | 311.77 | 899.92 | 1479.53 | 435.28 | 1338.23 | 2242.77 |
| | | Prosumer Balance | -653.96 | -2115.90 | -3267.64 | -1117.16 | -3895.78 | -6711.91 | -2115.90 | -5954.13 | -9782.04 |
| | | All Users Balance | -3680.37 | -11637.47 | -19566.74 | -3205.35 | -10451.02 | -17373.67 | -3348.93 | -8980.54 | -15313.67 |
| | PFET | Buyers Costs (P2P) | 397.36 | 1212.42 | 2060.22 | 392.90 | 1169.64 | 1915.56 | 244.82 | 665.44 | 1155.73 |
| | | Buyers Costs (Supp) | 2532.70 | 8017.15 | 13751.03 | 1563.27 | 5000.01 | 8132.78 | 854.05 | 2092.42 | 3877.51 |
| | | Buyers Total Costs | 2930.05 | 9229.56 | 15811.25 | 1956.17 | 6169.65 | 10048.34 | 1098.88 | 2757.85 | 5033.23 |
| | | Prosumers Costs | 801.66 | 2551.18 | 4016.18 | 1428.93 | 4795.70 | 8191.45 | 2551.18 | 7292.35 | 12024.81 |
| | | Prosumers Profits (P2P) | 397.36 | 1212.42 | 2060.22 | 392.90 | 1169.64 | 1915.56 | 244.82 | 665.44 | 1155.73 |
| | | Prosumers Profits (FiT) | 48.96 | 134.40 | 238.93 | 206.79 | 588.87 | 973.74 | 359.49 | 1151.43 | 1911.95 |
| | | Prosumers Balance | -355.34 | -1204.37 | -1717.03 | -829.24 | -3037.18 | -5302.15 | -1946.87 | -5475.49 | -8957.14 |
| | | All Users Balance | -3285.40 | -10433.93 | -17528.29 | -2785.41 | -9206.83 | -15350.49 | -3045.75 | -8233.34 | -13990.37 |
| | PP-LEM | Buyers Costs (P2P) | 353.31 | 993.71 | 1565.93 | 358.80 | 937.75 | 1475.15 | 251.42 | 551.33 | 937.97 |
| | | Buyers Costs (Supp) | 2532.70 | 8017.15 | 13751.03 | 1563.27 | 5000.01 | 8132.78 | 854.05 | 2092.42 | 3877.51 |
| | | Buyers Total Costs | 2886.00 | 9010.86 | 15316.96 | 1922.07 | 5937.76 | 9607.93 | 1105.47 | 2643.74 | 4815.47 |
| | | Prosumers Costs | 801.66 | 2551.18 | 4016.18 | 1428.93 | 4795.70 | 8191.45 | 2551.18 | 7292.35 | 12024.81 |
| | | Prosumers Profits (P2P) | 353.31 | 993.71 | 1565.93 | 358.80 | 937.75 | 1475.15 | 251.42 | 551.33 | 937.97 |
| | | Prosumers Profits (FiT) | 48.96 | 134.40 | 238.93 | 206.79 | 588.87 | 973.74 | 359.49 | 1151.43 | 1911.95 |
| | | Prosumers Balance | -399.39 | -1423.07 | -2211.33 | -863.34 | -3269.07 | -5742.56 | -1940.28 | -5589.60 | -9174.90 |
| | | All Users Balance | -3285.40 | -10433.93 | -17528.29 | -2785.41 | -9206.83 | -15350.49 | -3045.75 | -8233.34 | -13990.37 |
| With Battery | No P2P | Buyers Costs (Supp) | 3026.41 | 9521.57 | 16299.10 | 2088.19 | 6555.24 | 10661.75 | 1233.03 | 3026.41 | 5531.63 |
| | | Prosumers Costs (Supp) | 319.53 | 1122.19 | 1590.68 | 412.79 | 1838.90 | 3054.67 | 1122.19 | 2708.70 | 4257.53 |
| | | Prosumers Profits (FiT) | 44.60 | 128.81 | 227.62 | 93.11 | 265.98 | 387.21 | 128.81 | 360.98 | 590.90 |
| | | Prosumer Balance | -274.92 | -993.38 | -1363.06 | -319.68 | -1572.92 | -2667.46 | -993.38 | -2347.72 | -3666.63 |
| | | All Users Balance | -3301.34 | -10514.95 | -17662.16 | -2407.87 | -8128.16 | -13329.21 | -2226.41 | -5374.13 | -9198.26 |
| | PFET | Buyers Costs (P2P) | 531.79 | 1577.83 | 2725.76 | 799.91 | 2322.29 | 3807.07 | 597.28 | 1784.09 | 3115.28 |
| | | Buyers Costs (Supp) | 2342.97 | 7495.85 | 12817.50 | 959.69 | 3285.90 | 5384.19 | 209.81 | 396.97 | 870.20 |
| | | Buyers Total Costs | 2874.76 | 9073.68 | 15543.26 | 1759.60 | 5608.19 | 9191.27 | 807.09 | 2181.06 | 3985.47 |
| | | Prosumers Costs | 746.58 | 2400.50 | 3755.09 | 998.58 | 3565.45 | 6071.34 | 1452.47 | 3540.37 | 5837.25 |
| | | Prosumers Profits (P2P) | 531.79 | 1577.83 | 2725.76 | 799.91 | 2322.29 | 3807.07 | 597.28 | 1784.09 | 3115.28 |
| | | Prosumers Profits (FiT) | 0.00 | 0.00 | 0.00 | 0.00 | 0.00 | 0.00 | 3.47 | 34.13 | 35.71 |
| | | Prosumers Balance | -214.80 | -822.67 | -1029.33 | -198.67 | -1243.16 | -2264.27 | -851.72 | -1722.15 | -2686.26 |
| | | All Users Balance | -3089.55 | -9896.35 | -16572.58 | -1958.27 | -6851.35 | -11455.54 | -1658.81 | -3903.21 | -6671.74 |
| | PP-LEM | Buyers Costs (P2P) | 487.60 | 1335.06 | 2134.60 | 765.22 | 1944.51 | 3021.22 | 659.21 | 1488.88 | 2515.16 |
| | | Buyers Costs (Supp) | 2342.97 | 7495.85 | 12817.50 | 959.69 | 3285.90 | 5384.19 | 209.81 | 396.97 | 870.20 |
| | | Buyers Total Costs | 2830.57 | 8830.90 | 14952.09 | 1724.91 | 5230.41 | 8405.41 | 869.02 | 1885.85 | 3385.36 |
| | | Prosumers Costs | 746.58 | 2400.50 | 3755.09 | 998.58 | 3565.45 | 6071.34 | 1452.47 | 3540.37 | 5837.25 |
| | | Prosumers Profits (P2P) | 487.60 | 1335.06 | 2134.60 | 765.22 | 1944.51 | 3021.22 | 659.21 | 1488.88 | 2515.16 |
| | | Prosumers Profits (FiT) | 0.00 | 0.00 | 0.00 | 0.00 | 0.00 | 0.00 | 3.47 | 34.13 | 35.71 |
| | | Prosumers Balance | -258.98 | -1065.44 | -1620.49 | -233.36 | -1620.94 | -3050.12 | -789.79 | -2017.36 | -3286.37 |
| | | All Users Balance | -3089.55 | -9896.35 | -16572.58 | -1958.27 | -6851.35 | -11455.54 | -1658.81 | -3903.21 | -6671.74 |

**Table 14**
Profits and Costs within energy market in 2016.

| | | Ratio of Sellers | 25% | | | 50% | | | 75% | | |
|---|---|---|---|---|---|---|---|---|---|---|---|
| | | Number of Users | 40 | 120 | 200 | 40 | 120 | 200 | 40 | 120 | 200 |
| Without Battery | No P2P | Buyers Costs (Supp) | 32968.18 | 102131.99 | 176320.87 | 22551.30 | 70400.17 | 114702.02 | 13110.28 | 32968.18 | 59244.12 |
| | | Prosumers Costs (Supp) | 6717.19 | 21791.64 | 34394.06 | 12269.26 | 40972.44 | 69987.79 | 21791.64 | 62513.93 | 103094.00 |
| | | Prosumers Profits (FiT) | 3401.81 | 10127.71 | 17272.48 | 7073.49 | 20751.23 | 34254.74 | 10127.71 | 30929.02 | 51766.49 |
| | | Prosumer Balance | -3315.38 | -11663.93 | -17121.58 | -5195.77 | -20221.21 | -35733.05 | -11663.93 | -31584.90 | -51327.51 |
| | | All Users Balance | -36283.56 | -113795.92 | -193442.45 | -27747.07 | -90621.38 | -150435.07 | -24774.21 | -64553.08 | -110571.63 |
| | PFET | Buyers Costs (P2P) | 8116.65 | 24699.14 | 42091.50 | 6841.15 | 20658.43 | 33876.80 | 3885.98 | 10715.57 | 18568.04 |
| | | Buyers Costs (Supp) | 22518.59 | 70465.18 | 122181.66 | 13244.33 | 42525.14 | 69501.12 | 7288.09 | 18225.17 | 33287.02 |
| | | Buyers Total Costs | 30635.24 | 95164.32 | 164273.16 | 20085.48 | 63183.57 | 103377.92 | 11174.08 | 28940.74 | 51855.06 |
| | | Prosumers Costs | 6717.19 | 21791.64 | 34394.06 | 12269.26 | 40972.44 | 69987.79 | 21791.64 | 62513.93 | 103094.00 |
| | | Prosumers Profits (P2P) | 8116.65 | 24699.14 | 42091.50 | 6841.15 | 20658.43 | 33876.80 | 3885.98 | 10715.57 | 18568.04 |
| | | Prosumers Profits (FiT) | 1311.89 | 3794.35 | 6444.64 | 5212.09 | 15176.22 | 25214.56 | 8963.28 | 27980.42 | 46575.07 |
| | | Prosumers Balance | 2711.36 | 6701.85 | 14142.08 | -216.01 | -5137.79 | -10896.43 | -8942.39 | -23817.93 | -37950.89 |
| | | All Users Balance | -27923.89 | -88462.47 | -150131.08 | -20301.49 | -68321.35 | -114274.35 | -20116.46 | -52758.68 | -89805.95 |
| | PP-LEM | Buyers Costs (P2P) | 7435.95 | 20911.73 | 33241.04 | 6360.04 | 16772.05 | 26326.53 | 3859.01 | 8753.51 | 14746.50 |
| | | Buyers Costs (Supp) | 22518.59 | 70465.18 | 122181.66 | 13244.33 | 42525.14 | 69501.12 | 7288.09 | 18225.17 | 33287.02 |
| | | Buyers Total Costs | 29954.54 | 91376.91 | 155422.70 | 19604.37 | 59297.19 | 95827.65 | 11147.10 | 26978.68 | 48033.52 |
| | | Prosumers Costs | 6717.19 | 21791.64 | 34394.06 | 12269.26 | 40972.44 | 69987.79 | 21791.64 | 62513.93 | 103094.00 |
| | | Prosumers Profits (P2P) | 7435.95 | 20911.73 | 33241.04 | 6360.04 | 16772.05 | 26326.53 | 3859.01 | 8753.51 | 14746.50 |
| | | Prosumers Profits (FiT) | 1311.89 | 3794.35 | 6444.64 | 5212.09 | 15176.22 | 25214.56 | 8963.28 | 27980.42 | 46575.07 |
| | | Prosumers Balance | 2030.65 | 2914.44 | 5291.62 | -697.12 | -9024.16 | -18446.70 | -8969.36 | -25779.99 | -41772.43 |
| | | All Users Balance | -27923.89 | -88462.47 | -150131.08 | -20301.49 | -68321.35 | -114274.35 | -20116.46 | -52758.68 | -89805.95 |
| With Battery | No P2P | Buyers Costs (Supp) | 32968.18 | 102131.99 | 176320.87 | 22551.30 | 70400.17 | 114702.02 | 13110.28 | 32968.18 | 59244.12 |
| | | Prosumers Costs (Supp) | 1451.49 | 5115.70 | 6731.59 | 1667.90 | 7526.78 | 11597.87 | 5115.70 | 10641.86 | 15325.13 |
| | | Prosumers Profits (FiT) | 2337.04 | 6758.21 | 11685.01 | 4927.63 | 13994.68 | 22464.75 | 6758.21 | 20454.18 | 34042.33 |
| | | Prosumer Balance | 885.55 | 1642.51 | 4953.42 | 3259.73 | 6467.91 | 10866.88 | 1642.51 | 9812.32 | 18717.20 |
| | | All Users Balance | -32082.63 | -100489.48 | -171367.45 | -19291.57 | -63932.26 | -103835.14 | -11467.77 | -23155.86 | -40526.92 |
| | PFET | Buyers Costs (P2P) | 11683.30 | 35083.83 | 59853.09 | 13628.35 | 41278.18 | 67777.71 | 8021.81 | 22803.50 | 39970.04 |
| | | Buyers Costs (Supp) | 17332.56 | 55644.56 | 96700.25 | 3617.70 | 13021.84 | 21731.82 | 414.64 | 693.90 | 1709.19 |
| | | Buyers Total Costs | 29015.86 | 90728.39 | 156553.34 | 17246.05 | 54300.02 | 89509.53 | 8436.45 | 23497.40 | 41679.23 |
| | | Prosumers Costs | 5368.80 | 17742.06 | 27793.48 | 4458.80 | 16949.05 | 28377.18 | 6408.30 | 13802.79 | 21408.46 |
| | | Prosumers Profits (P2P) | 11683.30 | 35083.83 | 59853.09 | 13628.35 | 41278.18 | 67777.71 | 8021.81 | 22803.50 | 39970.04 |
| | | Prosumers Profits (FiT) | 5.01 | 20.31 | 28.24 | 1718.72 | 4467.14 | 7332.48 | 4484.10 | 14647.17 | 23784.63 |
| | | Prosumers Balance | 6319.50 | 17362.08 | 32087.85 | 10888.27 | 28796.26 | 46733.01 | 6097.61 | 23647.87 | 42346.21 |
| | | All Users Balance | -22696.35 | -73366.31 | -124465.49 | -6357.78 | -25503.76 | -42776.52 | -2338.84 | 150.48 | 666.98 |
| | PP-LEM | Buyers Costs (P2P) | 11085.98 | 30581.21 | 48820.32 | 12882.40 | 34138.51 | 53604.36 | 8287.44 | 18804.90 | 31947.59 |
| | | Buyers Costs (Supp) | 17332.56 | 55644.56 | 96700.25 | 3617.70 | 13021.84 | 21731.82 | 414.64 | 693.90 | 1709.19 |
| | | Buyers Total Costs | 28418.54 | 86225.77 | 145520.57 | 16500.10 | 47160.36 | 75336.18 | 8702.08 | 19498.80 | 33656.78 |
| | | Prosumers Costs | 5368.80 | 17742.06 | 27793.48 | 4458.80 | 16949.05 | 28377.18 | 6408.30 | 13802.79 | 21408.46 |
| | | Prosumers Profits (P2P) | 11085.98 | 30581.21 | 48820.32 | 12882.40 | 34138.51 | 53604.36 | 8287.44 | 18804.90 | 31947.59 |
| | | Prosumers Profits (FiT) | 5.01 | 20.31 | 28.24 | 1718.72 | 4467.14 | 7332.48 | 4484.10 | 14647.17 | 23784.63 |
| | | Prosumers Balance | 5722.18 | 12859.46 | 21055.08 | 10142.31 | 21656.60 | 32559.66 | 6363.24 | 19649.27 | 34323.76 |
| | | All Users Balance | -22696.35 | -73366.31 | -124465.49 | -6357.78 | -25503.76 | -42776.52 | -2338.84 | 150.48 | 666.98 |

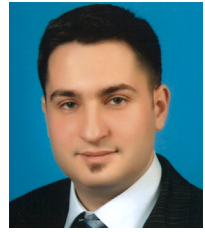

**Kamil Erdayandi** is pursuing a Ph.D. in Computer Science at The University of Manchester. He holds a B.Sc. degree in Electrical & Electronics Engineering with a dual degree in Computer Science, and earned an M.Sc. degree in Electronics. Keywords reflecting his current research interests encompass Data Privacy, Applied Cryptography, Game Theory, Homomorphic Encryption, Blockchain And Smart Grids.

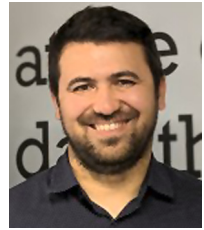

**Mustafa A. Mustafa** is a Senior Lecturer (Associate Professor) in the Department of Computer Science at The University of Manchester, where he leads the Trusted Digital Systems Cluster within the Centre for Digital Trust and Society. He earned his B.Sc. in communications from the Technical University of Varna in 2007, his M.Sc. in communications and signal processing from Newcastle University in 2010, and his Ph.D. in computer science from The University of Manchester in 2015. He was a post-doctoral research fellow with the imec-COSIC research group at KU Leuven. From July 2018 to June 2023, he was a Dame Kathleen Ollerenshaw Research Fellow at The University of Manchester.